\documentclass[12pt]{article}

\usepackage{graphicx} 
\usepackage{amsmath}

\usepackage{times}

\newenvironment{sciabstract}{%
\begin{quote} \bf}
{\end{quote}}

\title{Multistability and Intermediate Tipping of the Atlantic Ocean Circulation} 

\author
{Johannes Lohmann,$^{1\ast}$ Henk A. Dijkstra,$^{2}$ Markus Jochum,$^{1}$ \\Valerio Lucarini,$^{3}$ Peter D. Ditlevsen$^{1}$\\
\\
\normalsize{$^{1}$Physics of Ice, Climate and Earth, Niels Bohr Institute, University of Copenhagen, Denmark,}\\
\normalsize{$^{2}$Institute for Marine and Atmospheric research Utrecht, Utrecht University, Utrecht, the Netherlands}\\
\normalsize{$^{3}$Centre for the Mathematics of Planet Earth, University of Reading, Reading, UK}\\
\\
\normalsize{$^\ast$To whom correspondence should be addressed; E-mail: johannes.lohmann@nbi.ku.dk.}
}

\date{}

\begin{document} 

% Double-space the manuscript.

\baselineskip24pt

% Make the title.

\maketitle

% Place your abstract within the special {sciabstract} environment.

\begin{sciabstract}

  Tipping points (TP) in climate sub-systems are usually thought to occur at a well-defined, critical forcing parameter threshold, via destabilization of the system state by a single, dominant positive feedback. 
  However, coupling to other sub-systems, additional feedbacks, and spatial heterogeneity may promote further small-amplitude, abrupt reorganizations of geophysical flows at forcing levels lower than the critical threshold. Using a primitive-equation ocean model we simulate a collapse of the Atlantic Meridional Overturning Circulation (AMOC) due to increasing glacial melt. Considerably prior to the collapse, various abrupt, qualitative changes in AMOC variability occur. These intermediate tipping points (ITP) are transitions between multiple stable circulation states. Using 2.75 million years of model simulations, we uncover a very rugged stability landscape featuring parameter regions of up to nine coexisting stable states. The path to an AMOC collapse via a sequence of ITPs depends on the rate of change of the meltwater input. This challenges our ability to predict and define safe limits for TPs.
\end{sciabstract}

% In setting up this template for *Science* papers, we've used both
% the \section* command and the \paragraph* command for topical
% divisions.  Which you use will of course depend on the type of paper
% you're writing.  Review Articles tend to have displayed headings, for
% which \section* is more appropriate; Research Articles, when they have
% formal topical divisions at all, tend to signal them with bold text
% that runs into the paragraph, for which \paragraph* is the right
% choice.  Either way, use the asterisk (*) modifier, as shown, to
% suppress numbering.

\section*{Introduction}
The risk of climate tipping points (TP) due to increasing greenhouse warming has become an integral part of the scientific outreach about climate change to the general public. Nevertheless, they are usually assessed as low probability scenarios in commonly assumed pathways of future climate change \cite{IPCC21}. Indeed, the actual existence and anatomy of truly abrupt, singular TPs in real-world systems remains uncertain. 
Most state-of-the-art Earth system models (ESM) have difficulties simulating TP behavior, such as observed in the past climate record, perhaps due to stability biases \cite{VAL11,LIU17}. Their full stability landscape, i.e., the collection of changing equilibrium states as a function of boundary conditions (determined by atmospheric greenhouse gases), is computationally too expensive to assess at present.

Thus, TP research most often combines climate feedback assessments in observations and short ESM simulations with simpler, conceptual mathematical models, which in turn are used to interpret complex data sets. These models describe the state of a so-called tipping element (TE), which is a large-scale climate sub-system that is believed to possess a TP due to a dominant positive feedback. In the presence of non-linear feedbacks the models can display bistability, where in addition to a desired, present-day stable state (attractor), a second, undesired stable state exists for given boundary conditions. An important milestone in the development of this paradigm was due to Stommel \cite{STO61}, who proposed bistability and the existence of a reversed or collapsed circulation in a conceptual model of the AMOC. Another was the discovery by Budyko \cite{BUD68} and Sellers \cite{SEL69} of the existence - then confirmed by observational evidence \cite{HOF98} - of a snowball state that coexists with the warm state for given astronomical and astrophysical features of the Earth.
Since then, many parts of the Earth system have been proposed to display bistability between a present-day and a co-existing undesired state. This includes a savannah state of present-day rainforests, and an ice-free state of the present-day ice sheets \cite{LEN08}. 
For any such TE, the present-day state changes only sightly at first as the boundary conditions gradually deteriorate. But eventually a TP is reached where the present-day state loses stability, and the system is forced to undergo a transition to the undesired state as the only remaining stable solution. Prior to the TP, such conceptual models predict the existence of so-called early-warning signals (EWS), which are changes in the fluctuations of observables describing the TE state \cite{WIS84,HEL04,SCH09,SCH12}.

Consistent with this conceptual paradigm, a considerable number of large-scale climate sub-systems have been proposed to irreversibly and abruptly shift to an undesired state if the boundary conditions change sufficiently. At the same time, when staying below the purported TP no qualitative changes of the TE state are expected. In fact, many TEs have already been suggested to approach a TP, based on the analysis of EWS in historic and modern observations \cite{BOE21,BOE21a,MIC22,BOU22,LIU23,DIT23}. Conceptual models of TEs are without a doubt invaluable tools to understand complex real-world data sets and systems. Nevertheless, due to their coarse-grained nature, it is clear that they cannot describe the Earth system dynamics in full detail, and thus many researchers expect the dynamics and stability properties to be inevitably more complex. This suspected complexity is, however, difficult to show explicitly due to the high dimensionality and computational cost of climate models. In the following we list some mechanisms by which the stability landscape of a TE in the coupled climate system may become more complicated than typical conceptual models capture, and by which an eventual system collapse may not be a singular event that can be universally anticipated. 

In reality it may not always be warranted to model a TE in isolation, as its state may be substantially influenced by the state of another climate sub-system. If containing a positive feedback sensitive to the same control parameter, this other system may itself be a TE and display multiple stable states (multistability) and the possibility to tip. For instance, a bistable AMOC may interact with a bistable sub-polar gyre \cite{BOR13}. When tipping in one of the systems occurs before the tipping of the second system, the coupling leads to a qualitative, but potentially only subtle change in the state of the second TE. A model of the coupled system shows multistability beyond the bistability of the individual TEs. Coupling of several TEs has been considered in conceptual models of so-called cascades of TPs \cite{CAI16,DEK18,ROC18,WUN20,KLO20,WUN21,LOH21b}. However, even though there are examples of spontaneous or forced sequential regime shifts \cite{DRI13,ZHA14,KLE15}, this phenomenon has not been investigated systematically in high-dimensional ESMs.

Also in cases where a TE is not influenced substantially by other climate sub-systems
on the time scales of interest, simple bistable conceptual models may be insufficient to describe the dynamics.
First, individual TEs can be subject to more than one positive feedback due to different physical processes that influence the stability of the system. For instance, the positive salt-advection feedback of the AMOC may be complemented by localized positive convective feedbacks \cite{KUH07}. 
Second, even the action of only one dominant positive feedback on multiple weakly coupled domains in spatially extended, heterogeneous systems may lead to more complicated, spatially fragmented tipping \cite{VNE05,BEL12,RIE21,BAS22}. This aspect is not easily captured in conceptual models, but has been shown in studies of large-scale geophysical flows, which suggest the existence of multiple spatial patterns of convection that lead to different stable regimes of the AMOC \cite{LEN94,HUG94,RAH94,RAH95,SCH07,JOC12,KEA20}. 
Multistability beyond two coexisting states was also reported for snowball Earth transitions in intermediate-complexity ESMs \cite{LUC19,RAG22}, and was proposed to reflect the multiscale nature of the underlying quasipotential of the climate \cite{MAR21}. Finally, already relatively simple non-linear systems may intrinsically feature rich dynamics and much more pronounced multistability compared to the bistable dynamics of the usual TP paradigm \cite{FEU97,MAS98,PIK01,AST01,BAL05,FEU08}. 

Here we present a detailed investigation of the anatomy of a prominent TP, the collapse of the AMOC due to increasing glacial melt. This is done by careful construction of the stability landscape in a global ocean model, which exhibits many of the above aspects, such as spatial heterogeneity, high dimensionality, as well as a multitude of physical processes and scales. 
We investigate an ocean-only model with constant coupling to an atmospheric climatology. Such a choice enables us to simulate the model long enough to separate close-by but disconnected attractors, and it avoids the effectively stochastic atmosphere to mask the fine structure of the multistability in the ocean dynamics. In the real world, we expect that at some level close-by attractors can merge by atmospheric noise-induced transitions in the time scales of interest.
Using subtle changes in freshwater forcing, we find an abundance of abrupt qualitative AMOC changes that stem from a variety of multistable regimes with up to nine coexisting alternative attractors. Just like the TP associated with the collapse, the abrupt changes occur as one branch of attractors loses stability and the system undergoes a transition to another branch. For this phenomenon we use the term intermediate tipping points (ITP), since the changes occur prior to the AMOC collapse, and because the attractor branches are often confined to narrow parameter ranges. Unlike a full AMOC collapse, ITPs comprise only small changes in mean AMOC strength, but they can often be associated with specific changes in the spatio-temporal variability of the ocean circulation. Occurring well before the eventual collapse, they complicate the task of predicting a collapse from observational time series. 
We furthermore highlight that the complex landscape of coexisting basins of attraction permits rate-induced transitions, whereby the realized path to the AMOC collapse can be sensitive to initial conditions and the forcing variations. 

\section*{Results}

\subsection*{Hysteresis and intermediate tipping points}
\label{sec:hysteresis}

We start our analysis of the AMOC tipping by a hysteresis experiment, where the North Atlantic (NA) freshwater input is increased quasi-adiabatically at a rate of $1.5 \cdot 10^{-5}\, \text{Sv} /\text{yr}$ until AMOC collapse, and subsequently decreased again (Methods). Figure~1A shows the AMOC strength as a function of simulation time and freshwater forcing. Here and in the remainder of the paper, the AMOC strength is defined as the maximum value of the overturning streamfunction in the NA below 500m depth. A relatively narrow region of hysteresis is found (shaded area), where a collapsed AMOC state seems to coexist with a weakened, but still active circulation state with northern downwelling. The eventual collapse is not abrupt, but comprises three steps (inset in Fig.~1A) where the AMOC decrease temporarily slows down. 

The trajectory (blue) in Fig.~1A shows additional subtle, but discrete changes in the mean and variability. At moderate forcing, we see approximate quasi-periodic dynamics with a decadal-scale oscillation of close to 20 years. The spatio-temporal pattern of these oscillations is consistent with westward propagating Rossby waves in the NA (Fig.~S1), which were previously invoked as mechanism for multi-decadal Atlantic variability \cite{EDE01,GET05,DIJ05,FRA10}.
Mean and amplitude of the oscillations, as well as the relative strength of its periodic components, change abruptly in ITPs at several parameter values (e.g. Fig.~1B,C, and see Fig.~S2 for a wavelet analysis). At larger forcing, almost periodic oscillations with a period of roughly 11 years suddenly appear (Fig.~1D), which then abruptly die off leaving essentially chaotic low-amplitude variability with no dominant periodicities (see Fig.~1E and spectra in Fig.~S3). 

\subsection*{Stability landscape}
\label{sec:landscape}

The question arises whether the ITPs are abrupt changes in a single attractor of the system (so-called interior crisis of the attractor), or whether they reflect loss of stability of one attractor and a transitions to another (so-called boundary crisis of the attractor) \cite{OTT93}. Further, given that the overall collapse is gradual, it is not obvious whether the observed hysteresis is merely a transient. By performing equilibrium simulations that are branched off from the hysteresis simulation at various parameter values (red crosses in Fig.~1A, time series in Fig.~S4), we confirm that the hysteresis indeed reflects a coexistence of a vigorous and a collapsed AMOC state. The vigorous states in the hysteresis loop feature an already substantially weakened AMOC, but the overturning cell still spans from the polar NA down to the Southern Ocean (Fig.~2A,B). In addition, a partially-collapsed stable state with a further weakened AMOC exists, which is visited during the step-wise AMOC collapse of the hysteresis simulations. In this regime, there is a residual AMOC transport across the equator, but the overturning cell does not extend to the Southern Ocean (Fig.~2C). The maximum overturning strength in the NA is less than 7 Sv. In contrast, the collapsed states feature no equatorial transport by the AMOC (Fig.~2D), and its maximum strength in the NA is less than 3 Sv. 

There are indeed qualitative differences in the states obtained before and after the ITPs, as well as for the states found from the increasing and decreasing parameter sweep at the same forcing values (Fig.~S4). This shows that the ITPs are not transient, and that actual multistability exist.
To reveal the extent of the multistability regimes, we perform a continuation along the control parameter, using very small increments of the freshwater forcing and subsequent equilibration of the model. The continuation experiments comprise 280 simulations with an average duration of 7,005 years. Additional branches are found by large ensembles of simulations from independent initial conditions (Methods), yielding a total of 2.75 million simulation years. A bifurcation diagram showing all determined branches of stable states is given in Fig.~3A,B. For diagrams with other observables see Fig.~S5. 
In total, instead of just one pair of branches of states, corresponding to a vigorous and a collapsed AMOC, and two associated TPs, we identify 26 branches of attractors, as well as roughly 50 crises where an attractor loses stability. There are up to nine coexisting attractors for a fixed value of the freshwater forcing. Both the vigorous and collapsed regimes feature a number of coexisting attractors with only slightly differing mean AMOC strength. 

Figure~3C shows an ensemble of simulations with different initial conditions converging to the distinct states (see Fig.~S6 for ensembles at other parameter values). Time series of coexisting states at a selection of slices across the bifurcation diagram show that there are no trends in the AMOC strength, and that each state appears stationary (Fig.~3D-G). States of similar mean AMOC strength can differ substantially in their temporal variability (e.g. Fig.~3D), and some states feature rare, temporary excursions to lower values of the AMOC strength with multi-millennial recurrence times (e.g. red time series in Fig.~3E). The ITPs are caused by the loss of stability of one attractor and the subsequent transition to another attractor of different AMOC mean or variability, or both. One possible exception occurs at $F=0.283 \, \text{Sv}$, where the ITP rather seems to reflect an interior crisis of the attractor (Fig.~S7). 
The multistable cluster around $F=0.30 \, \text{Sv}$ to $F=0.32 \, \text{Sv}$, before the onset of the main hysteresis loop, causes the most clearly visible sequence of ITPs (Fig.~1C,D).

The edges of the main hysteresis loop are lined with short intervals of partially-collapsed AMOC states. They coexist with the vigorous or collapsed states, and form smaller-scale hysteresis loops with striking resemblance to the main hysteresis loop (see inset of Fig.~3A). As in the main loop, both upper and lower regimes of the smaller-scale loop comprise a cluster of multiple close-by attractors, which differ by only 0.1-0.2 Sv in AMOC strength. 
This suggests a hierarchical organization of the stability landscape, where clusters of attractors may appear at progressively smaller scales of the observable and the control parameter.

\subsection*{Spatio-temporal modes and heterogeneity}
\label{sec:symmetry}

The AMOC strength is tightly linked to the convective activity in different regions of the NA ocean. Since the bathymetry and atmospheric forcing are zonally asymmetric, transitions between convective and non-convective conditions might occur independently and at different values of the control parameter at the distinct NA locations.
As the freshwater forcing is increased, both the Labrador Sea and the Nordic Sea convective activity gradually weaken, as illustrated by the mean winter mixed layer depth (MLD) for a sequence of equilibrium simulations that comprise the states traversed during the transient AMOC collapse (Fig.~4). As the partially-collapsed AMOC state is reached, it appears that the Labrador Sea convection ceases while the Nordic Sea convection is still active (Fig.~4I).

Overall, we find that some of the co-existing stable states, and thus some of the ITPs, can be associated with changes in the convective pattern, while others cannot. 
Coexisting vigorous AMOC states at the same value of the control parameter show a qualitatively similar convective pattern. For instance, the convective activity in the Labrador and Nordic Seas is largely the same for the five coexisting stable states shown in Fig.~5A-E. However, there are substantial quantitative differences in MLD patterns, which are spatially correlated and distributed over the entire NA (Fig.~5K-N). 
States that are very close in mean AMOC strength (difference of 0.1-0.2 Sv) feature the same overall convective pattern and only relatively small quantitative differences in MLD (Fig.~S8). But also clusters that comprise both vigorous and partially-collapsed states can all share the same large-scale pattern (Fig.~S9a-f). On the other hand, the coexisting branches of collapsed AMOC states feature clearly distinct patterns of residual NA convection (Fig.~S9g-i), and some branches of partially-collapsed states feature different patterns compared to coexisting, vigorous branches (Fig.~S10a,d and S10b,e).

%%% Temporal variability of MLD.
The strength of vertical mixing in individual grid cells varies over time. The MLD shows considerable temporal variability that can be as large or larger than the differences in mean MLD between coexisting states (Fig.~5F-J). Usually, the large-scale convective pattern, i.e., the collection of grid cells that show deep mixing each winter, stays stationary over time. But there are states where a handful of neighboring grid cells oscillate between convective and non-convective winter conditions. For instance, the two coexisting states shown in Fig.~5A,B show decadal oscillations in cells south of Greenland and in the Nordic Seas (Fig.~S11 and Fig.~S12). Both states feature the exact same collection of cells that either participate in the oscillation or that are permanently convecting. As a result, the two states differ only in the period and amplitude of this oscillation and in the average mixing depth of cells. 

%%% TEMPORAL AMOC VARIABILITY.
%%% -> Conclusion for ITPs. Not necessarily dramatic shifts in patterns.
%%% shift in AMOC variability seems more striking!
%%% there should still be significant shifts in the ``quantitative'' pattern.
It is worth noting that the strength of NA MLD variability may not be directly reflected in the AMOC variability. The variability in MLD of states ON-1 and ON-2 (Fig.~5A,B) is roughly the same, but state ON-2 features a much larger AMOC variability (Fig.~5O). Similarly, states ON-4 and ON-5 feature a similar AMOC variability, but state ON-5 has much lower MLD variability (Fig.~5I,J). The AMOC variability in this cluster of states can also show different dominant frequencies (Fig.~S13). Indeed, coexisting states that are very close in mean AMOC strength (0.1-0.2 Sv) can differ substantially in their temporal AMOC variability and dominant periods (Fig.~S14) despite featuring near-identical convection patterns in the NA (Fig.~S8) and elsewhere.
As a result, we cannot directly link all ITPs with qualitative and abrupt changes of the convective pattern. But from the above analysis it is clear that due to the spatially distributed differences in MLD of the equilibrium states of different branches, and due to the striking differences in AMOC variability and spectrum, at least some ITPs manifest themselves as abrupt reorganizations in the spatio-temporal pattern. 

To test whether the previous findings are robust or a spurious effect due to the chosen resolution, we performed simulations with a doubled horizontal grid resolution, which again show ITPs prior to the AMOC collapse (Fig.~S15). While a direct confirmation of multistability similar to the lower resolution case has not been attempted due to high computational cost, the results suggest that also in the higher-resolution model there are coexisting states with subtly different AMOC strength, but qualitatively different variability. Branching off multi-millennial equilibrium simulations before and after ITPs reveals spatially distributed changes in MLD (Fig.~S16), and shows that at least one ITP can be associated with a qualitative change in the convective pattern (cessation of Nordic Sea convection in Fig.~S16b,d).

For both model resolutions it appears that some but not all coexisting states are associated with different spatial convection patterns. We next test more explicitly whether the ITPs are due to asymmetric boundary conditions and bathymetry in the NA. To this end we perform simulations with three alternative model configurations where the NA bathymetry and different boundary conditions are made zonally symmetric (see Methods and Fig.~S17). These all feature ITPs as in the fully heterogeneous case (Fig.~S18). The circulation and convective activity remain zonally asymmetric, and some ITPs (but not all) can be associated with qualitative changes in the convective pattern (Fig.~S18), where the main convective region quickly expands or migrates east-/westwards. The presence of ITPs and the underlying multistability thus seems not a consequence of heterogeneity in NA bathymetry and boundary conditions. Using more idealized, zonally symmetric model topography, future studies should investigate how heterogeneity outside of the NA, such as in the Southern Ocean, influences AMOC multistability.

\subsection*{Transient dynamics}
\label{sec:transient}

We now investigate in more detail how the transient dynamics during a linear parameter shift with non-negligible rate of change relate to the underlying stability landscape in the reference model setup. 
After initialization with a 10 kyr equilibrium simulation at $F=0.184 \, \text{Sv}$, we perform an ensemble of experiments where the freshwater forcing is ramped linearly from $F=0.184 \, \text{Sv}$ to $F=0.407 \, \text{Sv}$ at different rates of change (Fig.~S19). 
Figures~6A-C show three trajectories at increasing ramping speeds plotted on top of the bifurcation diagram of Fig.~3A.
We see that faster ramping leads to a weaker AMOC at a given parameter value. This may appear counter-intuitive: Since the equilibrium AMOC strength decreases with increasing freshwater forcing, one might expect the faster ramping trajectories to show a stronger instantaneous AMOC (lagging behind), thus with the trajectory curve above the equilibrium curve and not below as observed.

There is a similar sequence of ITPs for all but the fastest parameter shifts, where the time-dependent driving is very strong and the system is furthest away from its steady state properties (Fig.~S19). But, depending on the forcing rate, the AMOC variability after some ITPs is qualitatively different (Fig.~6A-C and Fig.~S20). Further, the ITPs occur at different forcing values. This is easily seen in Fig.~6A-C, where we marked the onset of the two sharpest ITPs (red and blue dashed lines). The parameter values where these occur are shown in Fig.~6D as a function of the rate. The value increases monotonically with decreasing forcing rate and slowly approaches the approximate critical value found in the equilibrium analysis (shaded bands in Fig.~6D).

Such tipping before the critical value that depends on the rate of change of the control parameter is referred to as rate-induced tipping \cite{ASH12}. Here, a slow rate of change below a critical rate does not lead to a tipping prior to the bifurcation point. This means that the curves in Fig.~6D should collapse onto the shaded bands for values smaller than the critical rate. In our simulation we cannot identify a critical rate, and it seems that the ITPs always occur before the critical value, also for very small rates of change where the corresponding state at higher AMOC strength is stable, and the tipping is not a transient effect (Fig.~S21). It may be that a critical value exists at even smaller rates, which we cannot test due to the computational constraints. Alternatively, this may be rate-induced tipping without a critical rate, as observed in simpler models of the AMOC \cite{LUC07,LOH21b}. In \cite{LOH21b} this was due to a non-smooth fold, whereby the basin boundary quickly comes very close to the moving attractor as the bifurcation is approached. This makes it impossible for the system to track the moving attractor, even for very small rates of change. In our multistable ocean model a similar situation may arise when multiple basins of attraction are tightly folded.

An additional transient effect is apparent from our simulations. For all but the slowest ramping speeds, the tipping to a state with stronger AMOC actually appears at a parameter value lower than the value where the corresponding branch of stable states begins to exist. For the branch associated with the first ITP this value is shown by the green line in Fig.~6A and D.
Either an additional coexisting attractor at higher AMOC strength has been missed, or the tipping is only transient. To investigate this, we start equilibrium simulations from the transient runs at a fixed forcing value, at which the simulations with fast ramping have already displayed an abrupt AMOC strengthening, but the slower ones have not (dotted line in Fig.~6D). At first, the faster simulations seem to converge to a stronger AMOC state (Fig.~7A and Fig.~S22). But eventually there is an abrupt transition back to a weaker AMOC state, which is identical to the one observed after slower ramping (Fig.~7B) and in the equilibrium analysis (gray trajectory). The trajectory with the fastest ramping shows an additional transient regime (Fig.~7C). This transient tipping could be understood as rate-induced tipping in an excitable system \cite{WIE23}. Since in our case there is also an underlying multistability, we speculate that it could also be related to the attraction by the stable manifold of a saddle, or to a ghost of the nearby branch of attractors \cite{IZH07}. This remains difficult to test in a model of this complexity. Similar attractions to saddles are known in rate-induced tipping as maximum canard trajectories \cite{WIE11}, but these are normally only expected for a narrow range of rates of change close to the critical rate. 

When instead starting constant-forcing simulations at a larger parameter value in the multistable regime, where some ensemble members have exhibited the second ITP discussed in Fig.~6, we find convergence to different attractors (Fig.~7D and Fig.~S23). Thus, different regions in phase space corresponding to different basins of attraction are visited, depending on the rate of parameter change during the traverse of the multistable regime. This leads to the rate-dependent differences in variability prior to the AMOC collapse discussed above.

\subsection*{Early-warning signals}
\label{sec:earlywarning}

One avenue for predicting an AMOC collapse is to use simulations with an ESM of high complexity to directly estimate the levels of freshwater forcing, global warming, or greenhouse gas concentrations at which a collapse occurs. Due to lack of processes and scales the model employed in our study is arguably not detailed enough for this task. In fact, the freshwater forcing at which the AMOC collapse occurs, as well as the number of ITPs observed prior to the collapse, depend sensitively on the model details. In a slightly changed model setup, where the relaxation time of the surface salinity to the forcing climatology is 5 instead of 2 years, we again see sharp ITPs, but in a sequence that is qualitatively different to what is obtained with 2-year relaxation time scale (Fig.~S24). 

Another avenue for TP prediction is to observe generic EWS from time series as an indication that a TP is near. Our model is suitable for testing the applicability of such EWS for an AMOC collapse in a high-dimensional model system, and in the presence of ITPs. EWS might be observable prior to each ITP, since each corresponds to a loss of stability of an attractor. But due to the a priori unknown number of ITPs before the AMOC collapse, it may not be possible to distinguish EWS prior to an arbitrary ITP from EWS prior to the AMOC collapse. EWS only measure changes in the local properties of the underlying dynamical system, and thus cannot tell whether the impending qualitative shift is a complete collapse or just a minor reorganization of the spatio-temporal climate pattern. 

A difficulty of actually observing EWS before any ITP is evident from the simulations with transient freshwater increase presented in the previous section. Due to rate-induced tipping, the ITPs do not necessarily occur at the parameter values where the corresponding attractors lose stability. In this case, prediction via critical slowing down is not possible in general \cite{LOH21b}. As an additional limitation to predictability, the critical rate beyond which a rate-induced tipping occurs can be fuzzy in chaotic systems \cite{LOH21,ASH21}. It is unknown how prevalent rate-dependent tipping is in multistable systems. But it seems plausible that the larger the number of coexisting and potentially folded basins of attraction, the harder it may be for a trajectory to stay exclusively in its current basin under a parameter shift, which is a condition for the non-existence of rate-induced tipping. 

Assuming EWS are consistently observed before ITPs, a continuous monitoring of the system during increasing freshwater forcing would lead to a jumpy behavior of EWS indicators, which repeatedly increase towards an ITP but then settle to a lower level as a different branch of attractors is reached. Additionally, there may be non-monotonic EWS trends between ITPs. Each branch of attractors loses stability in two ITPs, one for decreasing and one for increasing freshwater forcing. Thus, any universal EWS should increase towards either end of the branch, while being minimal at some mid-branch point. For each entire branch, the EWS should then be a u-shaped function of the control parameter (Fig.~8A,B). During a monotonic parameter shift, the system can enter a given branch at a forcing value before or after the EWS minimum. In the former case, there is an initial decrease of EWS before the increase associated with the next ITP (red trajectories). 

We next analyze the temporal variability of observables in our simulations of the equilibrium states along the control parameter, in order to investigate how an ideal EWS indicator may behave in the quasi-stationary limit of a slow, monotonic forcing increase.
We consider the sequence of attractor branches that would most likely be traversed during such a freshwater increase. The variance of the AMOC strength is chosen as EWS, since increased variance is a common EWS that may also be useful in some cases with oscillatory dynamics, such as Hopf bifurcations \cite{KEF13} and bifurcations of periodically driven systems \cite{WIL16}. 
Figure~8C,D shows jumps in variance at ITPs, and in some cases non-monotonic trends between ITPs. But u-shaped curves concerning the entire branches are not found. Most branches see decreasing variance, but this is not a robust feature since the direction of trends depend strongly on the observable used (compare Fig.~8C,D and Fig.~S25). 
The applicability of critical slowing down indicators (such as variance) in our model is thus not clear. The spectral changes of the AMOC strength leading up to the collapse do not indicate critical slowing down (Fig.~S3). Since usual EWS are not well suited for complex models featuring deterministic chaotic, as well as quasi-periodic dynamics, future studies should evaluate other statistics more suited for more complex, chaotic dynamics \cite{TAN18, GUT22}. 

\section*{Discussion}

Our results show that in a global ocean model an eventual collapse of the AMOC due to increased meltwater discharge is preceded by ITPs. These events are abrupt changes in the mean state and variability of the AMOC in response to slow and gradual meltwater increase. Occurring already far from the full AMOC collapse, they could be relevant in the context of global warming by impacting regional climates and other modes of climate variability. From a dynamical systems perspective ITPs are identical to what is commonly understood as a TP, i.e., the loss of stability of a branch of attractors and the subsequent transition to another coexisting attractor. By performing extensive equilibrium simulations, we showed that there is a large number of such attractors branches, i.e., a multitude of vigorous, collapsed and partially-collapsed stable states of the circulation that coexist for a given strength of freshwater forcing. The resulting rugged stability landscape features up to nine coexisting stable states and appears to be organized hierarchically since the attractors seem to come in clusters at different scales. As a result there may even be many more coexisting states on smaller scales of the observables and control parameter. 

Our findings are consistent with previous studies that identified 2-3 coexisting vigorous AMOC states in ocean models, and associated these with distinct spatial patterns of NA convection \cite{LEN94,HUG94,RAH94,RAH95,SCH07,KEA20}. 
Indeed we find coexisting states that have clearly different convective pattern. This may be understood as a spatially fragmented tipping of individual convection regions resulting from an underlying spatial heterogeneity, as has been proposed recently \cite{BAS22}. On the other hand, we uncover a large number of additional coexisting attractors that feature slightly different mean AMOC strength but near-identical NA convection patterns. Such states can also differ substantially in their temporal variability. 
By making the NA bathymetry and surface forcings zonally symmetric, we showed that gradually increasing freshwater forcing still induces abrupt transitions to different spatio-temporal patterns, but also abrupt, subtle changes in AMOC strength and temporal variability without changes in convective patterns. 

Extremely close-by coexisting states may be suspected an artifact of the model discretization and convective parameterizations, which may yield bistability between convective and non-convective states at individual grid cells, and thus a multitude of subtly different stable circulation patterns \cite{LEN94,LEN96,VEL98,DTO11}.
We cannot fully exclude this, as we lack a simplified model version with analytical solution \cite{DTO11}, and as the construction of the entire stability landscape of alternative model versions with different convective parameterizations is beyond the scope of the paper. Our model uses a smooth variant of convective adjustment (see Methods and the code in the Sec.~S1), which is expected to be less prone to spurious equilibria \cite{VEL98,DTO11}.
Further, our simulations strongly suggest differences in the dynamics beyond the grid cell level. We found coexisting pairs of states where there is no qualitative difference in convective activity on the grid cell level (Fig.~S11 and S12), i.e., there is no grid cell that is permanently convecting in one state, but permanently non-convecting in the other state. These states feature patches of neighboring grid cells that oscillate between convective and non-convective conditions at a decadal time scale. 
In general, there are marked quantitative differences in mean MLD of states that are very close-by in AMOC strength. These are typically distributed over the whole domain. Further, the differences are not static since there is a large internal temporal variability of MLD and AMOC strength. Amplitude and spectrum of this variability differ substantially for coexisting states. It should be tested whether close-by states persist at increased grid resolution. We have not demonstrated this explicitly due to computational constraints, but the existence of ITPs in our experiments with a doubled horizontal grid resolution at least suggests that multistability persists. 

Our usage of an ocean-only model excludes the possibility of the atmospheric temperature, precipitation, and wind fields to react to the freshwater-induced oceanic changes. Due to stabilizing/negative atmosphere-ocean feedbacks via a response of the atmospheric temperature field to the initial freshwater input, the AMOC weakening could be reduced, and thus a potential collapse may occur only at even larger forcing \cite{MAR22,MAR23}. Further, changes in wind fields are also important to the AMOC response to freshwater forcing, which can only be captured adequately when including a high-resolution atmospheric model \cite{MAR23}. It remains uncertain how atmospheric feedbacks influence a freshwater-induced AMOC decline. Positive atmospheric feedbacks have also been found when comparing the AMOC response in coupled climate models to ocean-only models \cite{STA11}. Further, it is not clear whether atmospheric feedbacks promote or prevent ITPs and high multistability. Negative feedbacks may remove some instabilities responsible for ITPs. On the other hand, further non-linear atmospheric processes could enable additional spatio-temporal shifts in different climate sub-systems, which could mean that ITPs become even more widespread.

While at present we cannot identify all mechanisms of multistability in our model, or prove that it is a robust feature, it is consistent with recent studies showcasing complicated stability landscapes of the Earth system \cite{LUC19,RAG22}, or complex ecosystems \cite{RIE21}. The potentially hierarchical organization of the stability diagram in Fig.~3A is in agreement with the recently proposed multiscale multistability of the climate \cite{MAR21}, and reminiscent of the self-similar bifurcation structure of recurring periodic windows at smaller and smaller scales in low-dimensional chaotic systems \cite{DEL21}. 
The resulting rugged stability landscape is illustrated schematically in Fig.~9. In such a landscape, small differences in the temporal variation of the control parameter could yield qualitatively different outcomes. This means, for instance, that safe limits for a temporary overshoot of the TP may be hard to define. 
Because multiple collapsed AMOC branches connect back to the vigorous regime (Fig.~3A), initial conditions and forcing trajectory could determine sensitively whether the AMOC can be recovered. Even a successful recovery after a reversal of the control parameter may lead to a different attractor than before the overshoot. This was seen in the hysteresis experiment, where a different equilibrium state was reached at a given control parameter after the decreasing parameter sweep (Fig.~S4).

The presence of high multistability adds another layer of fuzziness to the prediction of a TP, in addition to the indeterminacy of tipping for individual, rate-dependent TPs under chaotic dynamics \cite{LOH21,ASH21}. In a smooth landscape, impending transitions may be predicted by EWS if there is no early, rate-induced tipping (Fig.~9A). In a rugged landscape, however, predicting the eventual collapse is hard, since the sequence of impending ITPs may depend on initial conditions and the rate of the parameter shift. Depending on the rate of change, different basins of attraction are visited between ITPs. The resulting variety of different changes in variability that can precede the eventual AMOC collapse complicates the design of universal EWS. From the observation of EWS alone it is difficult to know how large the impending regime shift will be, since EWS only reflect changes in the local properties of the dynamical system. If they can be predicted - which may be hard since they occur before the parameter values where the corresponding attractors lose stability - ITPs may thus be mistaken for a complete AMOC collapse.

In summary, our simulations demonstrate an abundance of abrupt changes in spatio-temporal variability that can occur considerably prior to an AMOC collapse, and suggest a high degree of multistability of the ocean circulation. 
Verifying this in models of even higher dimensionality and complexity is computationally challenging. A complimentary approach may be detailed analyses of past AMOC transitions in climate proxy records. Indeed, many  Dansgaard-Oeschger cycles during the last glacial period, which are likely due to tipping between a collapsed and vigorous AMOC state, show abrupt, small-amplitude warming events prior to an abrupt NA cooling (i.e. an AMOC collapse) \cite{CAP10}. These so-called rebound events may be related to the ITPs reported here.

\section*{Methods}

\subsection*{Ocean model}
\label{sec:methods_model}

The simulations are done with Veros, a direct translation of the Fortran  backend of the ocean model PyOM2 \cite{EDE14} into Python/JAX \cite{HAE18,HAE21}. As primitive equation finite-difference global ocean model it has the same dynamical core and mixing physics as other z-level ocean models like MITgcm, or the CESM ocean model POP, but it is able to exploit GPU architectures and the ever expanding public Python libraries.
Mesoscale turbulence is represented using the Redi \cite{RED82} and Gent-McWilliams \cite{GEN90} parameterization for isopycnal and thickness diffusion (diffusivity of 1,000 m$^2$/s). We use the second order turbulence closure of Gaspar et al. \cite{GAS90} to account for diapycnal mixing. Here, a background diffusivity of 10$^{-5}$ m$^2$/s is employed, and static instabilities are removed by a gradual increase of the vertical diffusivity as the instability is approached. This corresponds to a smoothed variant of convective adjustment. The particular implementation in the Veros model is given in Section S1, and the resulting temporal evolution of the vertical diffusivity is shown in Fig.~S26 for an example simulation at some of the most convectively active grid cells.

The heat exchange boundary condition is expressed by a first-order Taylor expansion of the heat flux as a function of the anomaly of the modeled surface temperature with respect to a fixed surface temperature climatology \cite{BAR98}.  For this, we use ERA-40 climatologies of surface temperature and heat flux, as well as a climatology of the derivative of the heat flux with respect to changes in surface ocean temperature. The latter is derived from ERA-40 following Barnier et al. \cite{BAR95} and has a global yearly average 30.26 $WK^{-1} m^{-2}$.
Freshwater exchanges with the atmosphere are modeled by boundary conditions under which the sea surface salinity is relaxed towards a present-day climatological field within a given relaxation timescale.  By choosing a long timescale of 2 years, oceanic salinity anomalies are less efficiently damped by the atmospheric forcing compared to temperature anomalies. This enables the positive salt advection feedback, which can lead to AMOC multi-stability and tipping. 

The bathymetry is obtained by smoothing the ETOPO1 global relief model \cite{AMA09} with a Gaussian filter to match the grid resolution. The model domain ends at 80$^o$N and thereby does not feature an Arctic connection of ocean basins. The horizontal grid contains 90 longitudinal and 40 latitudinal cells, where the latitudinal resolution increases from 5.3$^o$ at the poles to 2.1$^o$ at the Equator, as well as 40 vertical layers, which increase in thickness from 23 m at the surface to 274 m at the bottom. To test the robustness of our results, we also performed simulations where the horizontal grid resolution is doubled.

Two previous studies investigated AMOC tipping in this ocean model, and further model details can be found therein \cite{LOH21,LOH22}. The latter study investigated tipping of the AMOC induced by volcanic cooling, and features the exact same model configuration as used here. The former study considered rate-induced tipping of the AMOC in a slightly different model setup.

\subsection*{Hysteresis experiment}
\label{sec:methods_hysteresis}

To perform a hysteresis experiment, a uniform freshwater forcing is introduced in the North Atlantic (NA) around Greenland (see Fig.~S27), which acts as the control parameter to induce an AMOC collapse. The model is initialized at the end of a 8000 year control run without freshwater forcing. Thereafter, a step-wise hysteresis experiment is performed. Here, the freshwater forcing is increased linearly by a value $\delta F=0.0043 \, \text{Sv}$ over a period of 200 years, after which it is held constant for 100 years. Thereafter, the next linear 200-year forcing increase by $\delta F$ and subsequent 100-year relaxation follows. This procedure is continued until $F=0.429 \, \text{Sv}$, after which the AMOC has fully collapsed. From the collapsed state, the forcing is reduced back to zero in the same step-wise manner. 

\subsection*{Continuation of attractor branches}
\label{sec:methods_continuation}

From the hysteresis experiment we start equilibrium simulations at various constant forcing values in order to evaluate whether the hysteresis reflects true multistability or merely transient dynamics. To this end, initial conditions are taken both from the increasing and decreasing parameter sweep of the hysteresis experiment. The several primary branch points for the equilibrium simulations are shown in Fig.~1A. They were chosen to correspond to forcing values just below and above those where a rapid change in AMOC state or variability was observed. 
After equilibration at each branch point, continuation experiments are performed as follows. 

From each branch point, two new simulations are started where the freshwater forcing is changed instantaneously by a very small positive or negative increment, respectively. Again, the forcing is held constant until the model has equilibrated, i.e., until the dynamics of relevant observables appear stationary. Here we require the AMOC strength to display no significant trend for at least 2,000 years.
Besides the AMOC strength, as observables we also evaluate the temporal evolution of surface and sub-surface temperatures, salinities and densities in different parts of the Atlantic.
The integration is done for at least 3,000 years, but mostly substantially longer. Longer simulation times are used either because the dynamics could not be deemed stationary after 3000 years, to assure long-term stability of the corresponding branch of attractors, or to better distinguish close-by states. 
From the two simulations with negative and positive forcing increments, we start new simulations with another increment in the same direction in order to continue the branch of attractors along the control parameter in both directions. For practical reasons, in many cases the new runs at altered forcing from a given simulation have been started somewhat before the simulation has fully equilibrated. While this still produces valid equilibrium simulations at the desired control parameter values, there is a slight risk of a premature escape from the current branch by a transition to a different basin of attraction due to rate-induced tipping. Continuation of a given branch is terminated when a qualitative shift to another, known branch of attractors, or to a previously unidentified state is observed. In the latter case, a new continuation experiment is started as above. 
Example timeseries obtained with this continuation method are given in Fig.~S28. 
We extend all branches in both directions by increasing and decreasing the control parameter until they are connected on either side with other known branches.
At this stage, the continuation procedure cannot yield any new attractors.
The method produced 280 distinct equilibrium simulations with a total simulation time of 1.96 million years, and an average simulation time of 7,005 years. About 50 percent of the simulations were longer than 5,300 years, and 28 percent longer than 10,000 years.
In addition, 119 simulations with perturbed initial conditions were performed, as explained in the following Section. 

As stability of an attractor branch is lost after a parameter increment, we sometimes find long, gradual trends and transient changes of variability before a new branch is reached (see Fig.~S29a-f for the most notable instances). In this case we simply increase the simulation time and integrate until the dynamics appears stationary and, if applicable, coincides with simulations on another branch. There were, however, a few cases of very long transients where we cannot establish with confidence what the stationary dynamics is (Fig.~S29g-i). As a result of these transients, some branches may in reality lose stability earlier or later than determined by our finite-time equilibrium simulations. 
The long average simulation times and the investigation of many branches with many realizations each makes it unlikely that entire branches are only of transient nature.
Nevertheless, the fact that long transients are possible makes it hard to prove that very close-by states are truly separate attractors, and to identify with full certainty at what parameter values branches merge.
In many cases, very long simulations of up to 25 kyr were performed to ensure that the attractors do not slowly merge over time (e.g. Fig.~S30). 
But this could not be done exhaustively, and from the ensemble simulations with perturbed initial conditions (see next Section) we found further candidates for distinct branches of close-by attractors. Because of computational time constraints, not all trajectories were integrated long enough to establish this. Thus, the close-by branches that have been continued and are shown in Fig.~3A (inset) may be interpreted as a representative sample of a cluster of close-by, potentially only marginally stable states. 

\subsection*{Perturbed initial condition ensembles}
\label{sec:methods_bisection}

The equilibrium simulations that were started from the hysteresis simulation and continued along the control parameter ultimately stem from a small set of initial conditions. The continuation method only yields one initial condition per parameter value on each branch.
To test whether a substantial number of branches may have been missed, and to show that the branches obtained during the continuation experiments are indeed attracting, we generate ensembles of additional initial conditions at a few specific values of the control parameter. For the values $F=0.339 \, \text{Sv}$, $F=0.343 \, \text{Sv}$, $F=0.347 \, \text{Sv}$, $F=0.351 \, \text{Sv}$, and $F=0.356 \, \text{Sv}$ within the multistable regime, we perform linear interpolations in between pairs of states on the vigorous and collapsed AMOC branches. The states were simply chosen as the snapshot at the last time step of the equilibrium simulations, and thus correspond to arbitrary samples from the attractor.

Then, 32 equally spaced initial conditions are chosen, which lie on a straight line in phase space, defined by the linear interpolation of these states. Each initial condition is integrated for at least 2,500 years, or until the dynamics were deemed stationary. The trajectories converged mostly to the attractors known from the continuation experiments, but in a few cases new branches were discovered and subsequently continued in parameter space. This ensemble consisted of 216 simulations, with a total simulation time of 792,000 years. Together with the continuation method, this yields a total simulation time of 2.75 million years.

\subsection*{Model experiments with zonally symmetric North Atlantic}
\label{sec:methods_symmetric}
To investigate the role of heterogeneity in the North Atlantic in generating the observed multistability, we consider three configurations with different degrees of symmetry in boundary conditions. In the first configuration, only the topography and bathymetry are altered. This is done by removing Greenland and Iceland, as well as filling in the Arctic sea and choosing straight meridional boundaries down to 45$^o$N at 300$^o$E and to 36$^o$N at 356$^o$E. A uniform depth of 4,000 meter is chosen, although the implementation of shelves in the model leads to depths of around 1,500 meters along the outer perimeter of the domain. In the second configuration, also the temperature and salinity forcing is made zonally symmetric, by zonal averaging in the North Atlantic. 
Finally, in the third configuration, also the horizontal wind stress fields are made zonally symmetric. Maps of the symmetrized bathymetry and forcing fields are shown in Fig.~S14. 
It is computationally prohibitive to construct the entire stability diagram for each case. Thus, we only perform one long transient simulation for each configuration, where the freshwater forcing is linearly ramped up very slowly until the AMOC collapses. This is sufficient to show whether ITPs occur. These simulations, started after 10 kyr of spinup at present-day forcing, are shown in Fig.~S18.

% Your references go at the end of the main text, and before the
% figures.  For this document we've used BibTeX, the .bib file
% scibib.bib, and the .bst file Science.bst.  The package scicite.sty
% was included to format the reference numbers according to *Science*
% style.

%BibTeX users: After compilation, comment out the following two lines and paste in
% the generated .bbl file. 
%\bibliographystyle{Science}
%\bibliography{refs}

\begin{thebibliography}{10}

\bibitem{IPCC21}
IPCC, {\it {Climate Change 2021: The Physical Science Basis. Contribution of
  Working Group I to the Sixth Assessment Report of the Intergovernmental Panel
  on Climate Change}\/} (Cambridge University Press, 2021).

\bibitem{VAL11}
P.~Valdes, {Built for stability}, {\it Nature Geosc\/} {\bf 4}, 414--416
  (2011).

\bibitem{LIU17}
W.~Liu, S.-P. Xie, Z.~Liu, J.~Zhu, {Overlooked possibility of a collapsed
  Atlantic Meridional Overturning Circulation in warming climate}, {\it Sci.
  Adv.\/} {\bf 3}, e1601666 (2017).

\bibitem{STO61}
H.~Stommel, {Thermohaline Convection with Two Stable Regimes of Flow}, {\it
  Tellus\/} {\bf 13}, 2 (1961).

\bibitem{BUD68}
M.~I. Budyko, {On the origin of glacial epochs}, {\it Meteorol. Gidrol\/} {\bf
  11}, 3--12 (1968).

\bibitem{SEL69}
W.~D. Sellers, {A climate model based on the energy balance of the
  earth-atmosphere system}, {\it J. Appl. Meteor.\/} {\bf 8}, 392--400 (1969).

\bibitem{HOF98}
P.~F. Hoffman, A.~J. Kaufman, G.~P. Halverson, D.~P. Schrag, {A Neoproterozoic
  Snowball Earth}, {\it Science\/} {\bf 281}, 1342--1346 (1998).

\bibitem{LEN08}
T.~M. Lenton, {\it et~al.\/}, {Tipping elements in the Earth’s climate
  system}, {\it PNAS\/} {\bf 105}, 1786--1793 (2008).

\bibitem{WIS84}
C.~Wissel, {A universal law of the characteristic return time near thresholds},
  {\it Oecologia\/} {\bf 65}, 101--107 (1984).

\bibitem{HEL04}
H.~Held, T.~Kleinen, {Detection of climate system bifurcations by degenerate
  fingerprinting}, {\it Geophys. Res. Lett.\/} {\bf 31}, L23207 (2004).

\bibitem{SCH09}
M.~Scheffer, {\it et~al.\/}, {Early-warning signals for critical transitions},
  {\it Nature\/} {\bf 461}, 53--59 (2009).

\bibitem{SCH12}
M.~Scheffer, {\it et~al.\/}, {Anticipating Critical Transitions}, {\it
  Science\/} {\bf 338}, 344--348 (2012).

\bibitem{BOE21}
N.~Boers, M.~Rypdal, {Critical slowing down suggests that the western Greenland
  Ice Sheet is close to a tipping point}, {\it PNAS\/} {\bf 21}, e2024192118
  (2021).

\bibitem{BOE21a}
N.~Boers, {Observation-based early-warning signals for a collapse of the
  Atlantic Meridional Overturning Circulation}, {\it Nature Clim. Change\/}
  {\bf 11}, 680--688 (2021).

\bibitem{MIC22}
S.~L.~L. Michel, {\it et~al.\/}, {Early warning signal for a tipping point
  suggested by a millennial Atlantic Multidecadal Variability reconstruction},
  {\it Nature Comm.\/} {\bf 13}, 5176 (2022).

\bibitem{BOU22}
C.~Boulton, T.~M. Lenton, N.~Boers, {Pronounced loss of Amazon rainforest
  resilience since the early 2000s}, {\it Nature Clim. Change\/} {\bf 12},
  271--278 (2022).

\bibitem{LIU23}
T.~Liu, {\it et~al.\/}, {Teleconnections among tipping elements in the Earth
  system}, {\it Nature Clim. Change\/} {\bf 13}, 67--74 (2023).

\bibitem{DIT23}
P.~Ditlevsen, S.~Ditlevsen, {Warning of a forthcoming collapse of the Atlantic
  meridional overturning circulation}, {\it Nature Comm.\/} {\bf 14}, 4254
  (2023).

\bibitem{BOR13}
A.~Born, T.~F. Stocker, C.~C. Raible, A.~Levermann, {Is the Atlantic subpolar
  gyre bistable in comprehensive coupled climate models?}, {\it Clim Dyn\/}
  {\bf 40}, 2993--3007 (2013).

\bibitem{CAI16}
Y.~Cai, T.~M. Lenton, T.~S. Lontzek, {Risk of multiple interacting tipping
  points should encourage rapid CO$_2$ emission reduction}, {\it Nature Clim.
  Change\/} {\bf 6}, 520-525 (2016).

\bibitem{DEK18}
M.~M. Dekker, A.~S. von~der Heydt, H.~A. Dijkstra, {Cascading transitions in
  the climate system}, {\it Earth Syst. Dynam.\/} {\bf 9}, 1243--1260 (2018).

\bibitem{ROC18}
J.~C. Rocha, G.~Peterson, S.~Bodin, {\"{O}}and~Levin, {Cascading regime shifts
  within and across scales}, {\it Science\/} {\bf 362}, 1379--1383 (2018).

\bibitem{WUN20}
N.~Wunderling, M.~Gelbrecht, R.~Winkelmann, J.~Kurths, J.~F. Donges, {Basin
  stability and limit cycles in a conceptual model for climate tipping
  cascades}, {\it New J. Phys.\/} {\bf 22}, 123031 (2020).

\bibitem{KLO20}
A.~K. Klose, V.~Karle, R.~Winkelmann, J.~F. Donges, {Emergence of cascading
  dynamics in interacting tipping elements of ecology and climate}, {\it R.
  Soc. Open Sci.\/} {\bf 7}, 200599 (2020).

\bibitem{WUN21}
N.~Wunderling, J.~F. Donges, J.~Kurths, R.~Winkelmann, {Interacting tipping
  elements increase risk of climate domino effects under global warming}, {\it
  Earth Syst. Dynam.\/} {\bf 12}, 601--619 (2021).

\bibitem{LOH21b}
J.~Lohmann, D.~Castellana, P.~D. Ditlevsen, H.~A. Dijkstra, {Abrupt climate
  change as a rate-dependent cascading tipping point}, {\it Earth Syst.
  Dynam.\/} {\bf 12}, 819--835 (2021).

\bibitem{DRI13}
S.~Drijfhout, E.~Gleeson, H.~A. Dijkstra, V.~Livina, {Spontaneous abrupt
  climate change due to an atmospheric blocking--sea-ice--ocean feedback in an
  unforced climate model simulation}, {\it PNAS\/} {\bf 110}, 19713--19718
  (2013).

\bibitem{ZHA14}
X.~Zhang, G.~Lohmann, G.~Knorr, C.~Purcell, {Abrupt glacial climate shifts
  controlled by ice sheet changes}, {\it Nature\/} {\bf 512}, 290--294 (2014).

\bibitem{KLE15}
H.~Kleppin, M.~Jochum, B.~Otto-Bliesner, C.~A. Shields, S.~Yeager, {Stochastic
  Atmospheric Forcing as a Cause of Greenland Climate Transitions}, {\it J.
  Climate\/} {\bf 28}, 7741--7763 (2015).

\bibitem{KUH07}
T.~Kuhlbrodt, {\it et~al.\/}, {On the driving processes of the Atlantic
  meridional overturning circulation}, {\it Rev. Geophys.\/} {\bf 45}, RG2001
  (2007).

\bibitem{VNE05}
E.~van Nes, M.~Scheffer, {Implications of spatial heterogeneity for
  catastrophic regime shifts in ecosystems}, {\it Ecology\/} {\bf 86},
  1797--1807 (2005).

\bibitem{BEL12}
G.~Bel, A.~Hagberg, E.~Meron, {Gradual regime shifts in spatially extended
  ecosystems}, {\it Theor Ecol\/} {\bf 5}, 591--604 (2012).

\bibitem{RIE21}
M.~Rietkerk, {\it et~al.\/}, {Evasion of tipping in complex systems through
  spatial pattern formation}, {\it Science\/} {\bf 374}, 169 (2021).

\bibitem{BAS22}
R.~Bastiaansen, H.~A. Dijkstra, A.~von~der Heydt, {Fragmented tipping in a
  spatially heterogeneous world}, {\it Environ. Res. Lett.\/} {\bf 17}, 045006
  (2022).

\bibitem{LEN94}
G.~Lenderink, R.~J. Haarsma, {Variability and multiple equilibria of the
  thermohaline circulation associated with deep-water formation}, {\it J. Phys.
  Oceanogr.\/} {\bf 24}, 1480--1493 (1994).

\bibitem{HUG94}
T.~M.~C. Hughes, A.~J. Weaver, {Multiple Equilibria of an Asymmetric Two-Basin
  Ocean Model}, {\it J Phys Oceanogr\/} {\bf 24}, 619--637 (1994).

\bibitem{RAH94}
S.~Rahmstorf, {Rapid climate transitions in a coupled ocean-atmosphere model},
  {\it Nature\/} {\bf 372}, 82--84 (1994).

\bibitem{RAH95}
S.~Rahmstorf, {Multiple convection patterns and thermohaline flow in an
  idealized OGCM}, {\it J Clim\/} {\bf 8}, 3028--3039 (1995).

\bibitem{SCH07}
M.~Schulz, A.~Prange, A.~Klocker, {Low-frequency oscillations of the Atlantic
  Ocean meridional overturning circulation in a coupled climate model}, {\it
  Clim. Past\/} {\bf 3}, 97--107 (2007).

\bibitem{JOC12}
M.~Jochum, {\it et~al.\/}, {True to Milankovitch: Glacial Inception in the New
  Community Climate System Model}, {\it J. Climate\/} {\bf 25}, 2226--2239
  (2012).

\bibitem{KEA20}
A.~Keane, B.~Krauskopf, T.~M. Lenton, {Signatures Consistent with
  Multifrequency Tipping in the Atlantic Meridional Overturning Circulation},
  {\it Phys Rev Lett\/} {\bf 125}, 228701 (2020).

\bibitem{LUC19}
V.~Lucarini, T.~Bódai, {Transitions across Melancholia States in a Climate
  Model: Reconciling the Deterministic and Stochastic Points of View}, {\it
  Phys Rev Lett\/} {\bf 122}, 158701 (2019).

\bibitem{RAG22}
C.~Ragon, {\it et~al.\/}, {Robustness of Competing Climatic States}, {\it J
  Clim\/} {\bf 35}, 2769 (2022).

\bibitem{MAR21}
G.~Margazoglou, T.~Grafke, A.~Laio, V.~Lucarini, {Dynamical landscape and
  multistability of a climate model}, {\it Proc. R. Soc. A\/} {\bf 477},
  20210019 (2021).

\bibitem{FEU97}
U.~Feudel, C.~Grebogi, {Multistability and the control of complexity}, {\it
  Chaos\/} {\bf 7}, 597 (1997).

\bibitem{MAS98}
C.~Masoller, N.~B. Abraham, {Stability and dynamical properties of the
  coexisting attractors of an external-cavity semiconductor laser}, {\it Phys.
  Rev. A\/} {\bf 57}, 1313--1322 (1998).

\bibitem{PIK01}
A.~Pikovsky, O.~Popovych, Y.~Maistrenko, {Resolving Clusters in Chaotic
  Ensembles of Globally Coupled Identical Oscillators}, {\it Phys. Rev.
  Lett.\/} {\bf 87}, 044102 (2001).

\bibitem{AST01}
V.~Astakhov, A.~Shabunin, W.~Uhm, S.~Kim, {Multistability formation and
  synchronization loss in coupled Hénon maps: Two sides of the single
  bifurcational mechanism}, {\it Phys. Rev. E\/} {\bf 63}, 056212 (2001).

\bibitem{BAL05}
A.~G. Balanov, N.~B. Janson, E.~Schöll, {Delayed feedback control of chaos:
  Bifurcation analysis}, {\it Phys. Rev. E\/} {\bf 71}, 016222 (2005).

\bibitem{FEU08}
U.~Feudel, {Complex dynamics in multistable systems}, {\it Int J Bif Chaos\/}
  {\bf 18}, 1607--1626 (2008).

\bibitem{EDE01}
C.~Eden, J.~Willebrand, {Mechanism of Interannual to Decadal Variability of the
  North Atlantic Circulation}, {\it J. Climate\/} {\bf 14}, 2266--2280 (2001).

\bibitem{GET05}
J.~Getzlaff, C.~W. Böning, C.~Eden, A.~Biastoch, {Signal propagation related
  to the North Atlantic overturning}, {\it Geophys. Res. Lett.\/} {\bf 32},
  L09602 (2005).

\bibitem{DIJ05}
H.~A. Dijkstra, L.~te~Raa, M.~Schmeits, J.~Gerrits, {On the physics of the
  Atlantic Multidecadal Oscillation}, {\it Ocean Dyn\/} {\bf 56}, 36--50
  (2006).

\bibitem{FRA10}
L.~M. Frankcombe, A.~von~der Heydt, H.~A. Dijkstra, {North Atlantic
  Multidecadal Climate Variability: An Investigation of Dominant Time Scales
  and Processes}, {\it J Climate\/} {\bf 23}, 3626--3638 (2010).

\bibitem{OTT93}
E.~Ott (Cambridge University Press, 1993).

\bibitem{ASH12}
P.~Ashwin, S.~Wieczorek, R.~Vitolo, P.~Cox, {Tipping points in open systems:
  bifurcation, noise-induced and rate-dependent examples in the climate
  system}, {\it Phil. Trans. R. Soc. A\/} {\bf 370}, 1166--1184 (2012).

\bibitem{LUC07}
V.~Lucarini, S.~Calmanti, V.~Artale, {Experimental Mathematics: Dependence of
  the Stability Properties of a Two-Dimensional Model of the Atlantic Ocean
  Circulation on the Boundary Conditions}, {\it Russ. Journal of Math. Phys.\/}
  {\bf 14}, 224--231 (2007).

\bibitem{WIE23}
S.~Wieczorek, C.~Xie, P.~Ashwin, {Rate-induced tipping: thresholds, edge states
  and connecting orbits}, {\it Nonlinearity\/} {\bf 36}, 3238 (2023).

\bibitem{IZH07}
E.~M. Izhikevich (MIT Press, 2007).

\bibitem{WIE11}
S.~Wieczorek, P.~Ashwin, C.~M. Luke, P.~M. Cox, {Excitability in ramped
  systems: the compost-bomb instability}, {\it Proc. R. Soc. A\/} {\bf 467},
  1243--1269 (2011).

\bibitem{LOH21}
J.~Lohmann, P.~D. Ditlevsen, {Risk of tipping the overturning circulation due
  to increasing rates of ice melt}, {\it PNAS\/} {\bf 118}, e2017989118 (2021).

\bibitem{ASH21}
P.~Ashwin, J.~Newman, {Physical invariant measures and tipping probabilities
  for chaotic attractors of asymptotically autonomous systems}, {\it Eur. Phys.
  J. Spec. Top.\/} {\bf 230}, 3235--3248 (2021).

\bibitem{KEF13}
S.~Kéfi, V.~Dakos, M.~Scheffer, E.~H. van Nes, M.~Rietkerk, {Early warning
  signals also precede non-catastrophic transitions}, {\it Oikos\/} {\bf 122},
  641--648 (2013).

\bibitem{WIL16}
M.~S. Williamson, S.~Bathiany, T.~M. Lenton, {Early warning signals of tipping
  points in periodically forced systems}, {\it Earth Syst. Dynam.\/} {\bf 7},
  313-326 (2016).

\bibitem{TAN18}
A.~Tantet, V.~Lucarini, F.~Lunkeit, H.~A. Dijkstra, {Crisis of the chaotic
  attractor of a climate model: a transfer operator approach}, {\it
  Nonlinearity\/} {\bf 226}, 2221--2251 (2017).

\bibitem{GUT22}
M.~S. Gutiérrez, V.~Lucarini, {On some aspects of the response to stochastic
  and deterministic forcings}, {\it J. Phys. A: Math. Theor.\/} {\bf 55},
  425002 (2022).

\bibitem{LEN96}
G.~Lenderink, R.~J. Haarsma, {apid convective transitions in the presence of
  sea ice}, {\it J. Phys. Oceanogr.\/} {\bf 26}, 1448--1467 (1996).

\bibitem{VEL98}
M.~Vellinga, {Multiple Equilibria in Ocean Models as a Side Effect of
  Convective Adjustment}, {\it J Phys Oceanogr\/} {\bf 28}, 621--633 (1998).

\bibitem{DTO11}
M.~den Toom, H.~A. Dijkstra, F.~W. Wubs, {Spurious multiple equilibria
  introduced by convective adjustment}, {\it Ocean Modelling\/} {\bf 38},
  126--137 (2011).

\bibitem{MAR22}
T.~Martin, A.~Biastoch, G.~Lohmann, U.~Mikolajewicz, X.~Wang, {On timescales
  and reversibility of the ocean's response to enhanced Greenland Ice Sheet
  melting in comprehensive climate models}, {\it Geophys. Res. Lett.\/} {\bf
  49}, e2021GL097114 (2022).

\bibitem{MAR23}
T.~Martin, A.~Biastoch, {On the ocean's response to enhanced Greenland runoff
  in model experiments: relevance of mesoscale dynamics and atmospheric
  coupling}, {\it Ocean Sci.\/} {\bf 19}, 141--167 (2023).

\bibitem{STA11}
D.~Stammer, N.~Agarwal, P.~Herrmann, A.~Köhl, C.~R. Mechoso, {Response of a
  Coupled Ocean–Atmosphere Model to Greenland Ice Melting}, {\it Surveys in
  Geophysics\/} {\bf 32}, 621--642 (2011).

\bibitem{DEL21}
R.~de~Leo, J.~A. Yorke, {Infinite towers in the graphs of many dynamical
  systems}, {\it Nonlinear Dyn\/} {\bf 105}, 813--835 (2021).

\bibitem{CAP10}
E.~Capron, {\it et~al.\/}, {Millennial and sub-millennial scale climatic
  variations recorded in polar ice cores over the last glacial period}, {\it
  Clim. Past\/} {\bf 6}, 345--365 (2010).

\bibitem{EDE14}
C.~Eden, {pyOM2.0 Documentation, available at:
  https://wiki.cen.uni-hamburg.de/ifm/TO/pyOM2} pp. pyOM2.0 Documentation,
  available at: https://wiki.cen.uni--hamburg.de/ifm/TO/pyOM2 (2014).

\bibitem{HAE18}
D.~H{\"a}fner, {\it et~al.\/}, {Veros v0.1 -- a fast and versatile ocean
  simulator in pure Python}, {\it Geosci. Model Dev.\/} {\bf 11}, 3299--3312
  (2018).

\bibitem{HAE21}
D.~H{\"a}fner, R.~Nuterman, M.~Jochum, {Fast, Cheap, and Turbulent—Global
  Ocean Modeling With GPU Acceleration in Python}, {\it JAMES\/} {\bf 13},
  e2021MS002717 (2021).

\bibitem{RED82}
M.~H. Redi, {Oceanic isopycnal mixing by coordinate rotation}, {\it J. Phys.
  Oceanogr.\/} {\bf 12}, 1154--1158 (1982).

\bibitem{GEN90}
P.~R. Gent, J.~C. McWilliams, {Isopycnal mixing in ocean circulation models},
  {\it J. Phys. Oceanogr.\/} {\bf 20}, 150--155 (1990).

\bibitem{GAS90}
P.~Gaspar, Y.~Grégoris, J.~M. Lefevre, {A simple Eddy kinetic energy model for
  simulations of the oceanic vertical mixing: Tests at station Papa and
  long-term upper ocean study}, {\it J. Geophys. Res.\/} {\bf 95}, 16179--16193
  (1990).

\bibitem{BAR98}
B.~Barnier, {\it {Forcing the oceans}\/} (Springer, 1998), pp. 45--80.

\bibitem{BAR95}
B.~Barnier, L.~Siefridt, P.~Marchesiello, {Thermal forcing for a global ocean
  circulation model using a three-year climatology of ECMWF analyses}, {\it J.
  Marine Syst.\/} {\bf 6}, 363--380 (1995).

\bibitem{AMA09}
C.~Amante, B.~W. Eakins, {ETOPO1 1 arc-minute global relief model: procedures,
  data sources and analysis}, {\it Tech. rep.\/} (2009).

\bibitem{LOH22}
J.~Lohmann, A.~Svensson, {Ice core evidence for major volcanic eruptions at the
  onset of Dansgaard-Oeschger warming events }, {\it Clim. Past.\/} {\bf 18},
  2021--2043 (2022).

\end{thebibliography}

\section*{Acknowledgments}
{\bf Acknowledgments:} We thank Roman Nuterman and the Danish Center for Climate Computing for supporting the simulations with the Veros ocean model, and Ulrike Feudel, Peter Ashwin, and Andrew Keane for valuable discussions.
{\bf Funding: } The project has received funding from the European Union’s Horizon 2020 research and innovation programme under grant agreement No. 820970, and from Danmarks
Frie Forskningsfond under grant No. 2032-00346B.
{\bf Author contributions: } JL designed and performed the research. The paper was written by JL with input from all co-authors. All authors discussed and interpreted the results.
{\bf Competing interests: } The authors declare no competing interest.
{\bf Data and materials availability: } The raw model simulation data that underlie the findings of this study are deposited in the Electronic Research Data Archive repository of the University of Copenhagen and can be accessed at https://sid.erda.dk/sharelink/HpsKecpv8G.

\section*{Supplementary materials}
%Materials and Methods\\
Section S1, and Figs. S1 to S30\\

% For your review copy (i.e., the file you initially send in for
% evaluation), you can use the {figure} environment and the
% \includegraphics command to stream your figures into the text, placing
% all figures at the end.  For the final, revised manuscript for
% acceptance and production, however, PostScript or other graphics
% should not be streamed into your compliled file.  Instead, set
% captions as simple paragraphs (with a \noindent tag), setting them
% off from the rest of the text with a \clearpage as shown  below, and
% submit figures as separate files according to the Art Department's
% instructions.

%\clearpage

%\noindent {\bf Fig. 1.} Please do not use figure environments to set
%up your figures in the final (post-peer-review) draft, do not include graphics in your
%source code, and do not cite figures in the text using \LaTeX\
%\verb+\ref+ commands.  Instead, simply refer to the figure numbers in
%the text per {\it Science\/} style, and include the list of captions at
%the end of the document, coded as ordinary paragraphs as shown in the
%\texttt{scifile.tex} template file.  Your actual figure files should
%be submitted separately.

\begin{figure}%[floatfix]%!htb
\includegraphics[width=0.99\textwidth]{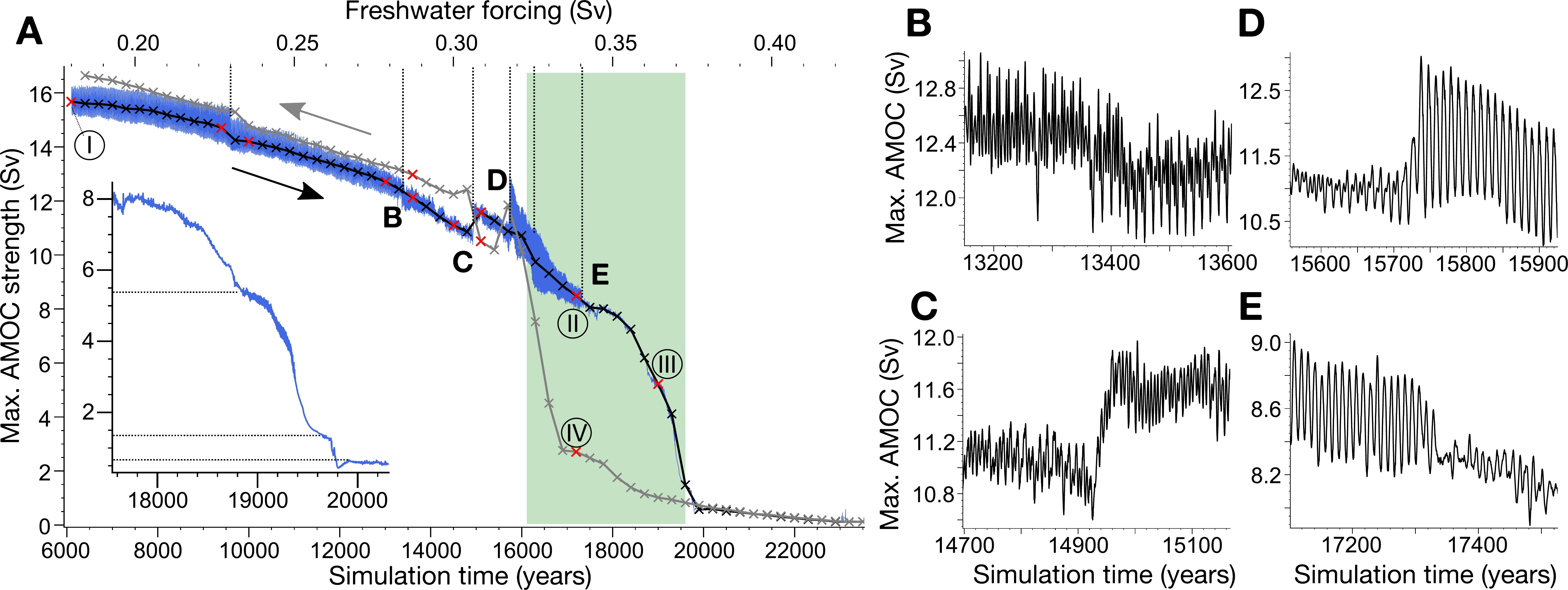}
\caption{\label{fig:hysteresis} 
{\bf Intermediate tipping points prior to a freshwater-induced AMOC collapse.}
{\bf A} Maximum AMOC strength as a function of the freshwater forcing during a hysteresis experiment (Methods). The black (gray) crosses are 50-year averages obtained after each forcing increment during the increasing (decreasing) part of the hysteresis experiment. In blue we show the actual trajectory in 5-year averages during the increasing parameter sweep. At the red crosses we start equilibrium simulations with fixed forcing, which form the basis for the construction of the stability landscape (Methods). The inset is a close-up of the collapsing trajectory, which seems to consist of 3 steps. 
{\bf B-E} Close-ups of the trajectory in 1-year averages showing discontinuous changes in mean and variability of the AMOC strength, as a response to the gradual change in forcing.
}
\end{figure}

\begin{figure}%[floatfix]%!htb
\includegraphics[width=0.8\textwidth]{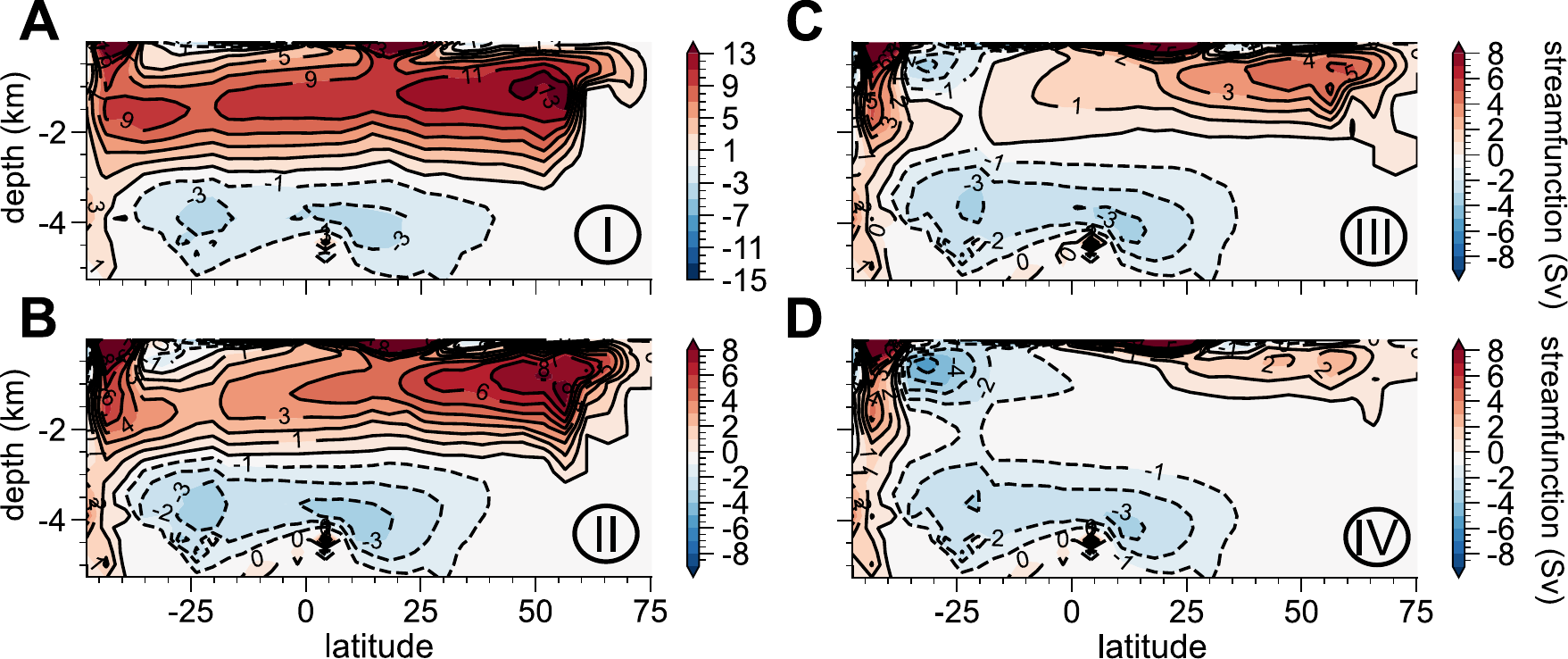}
\caption{\label{fig:streamfunctions} 
{\bf Overturning streamfunctions of different circulation regimes.} Meridional overturning streamfunctions obtained from 100-year averages of the velocity fields in the Atlantic for equilibrium simulations that were started from the hysteresis simulation at the points marked by roman numerals in Fig.~1A. Shown are states corresponding to a vigorous AMOC for moderate freshwater forcing (I), the vigorous regime close to the collapse (II), a partially-collapsed state (III), and a collapsed state (IV). Note the different color bar scale for panel I. 
}
\end{figure}

\begin{figure}%[floatfix]%!htb
\includegraphics[width=0.99\textwidth]{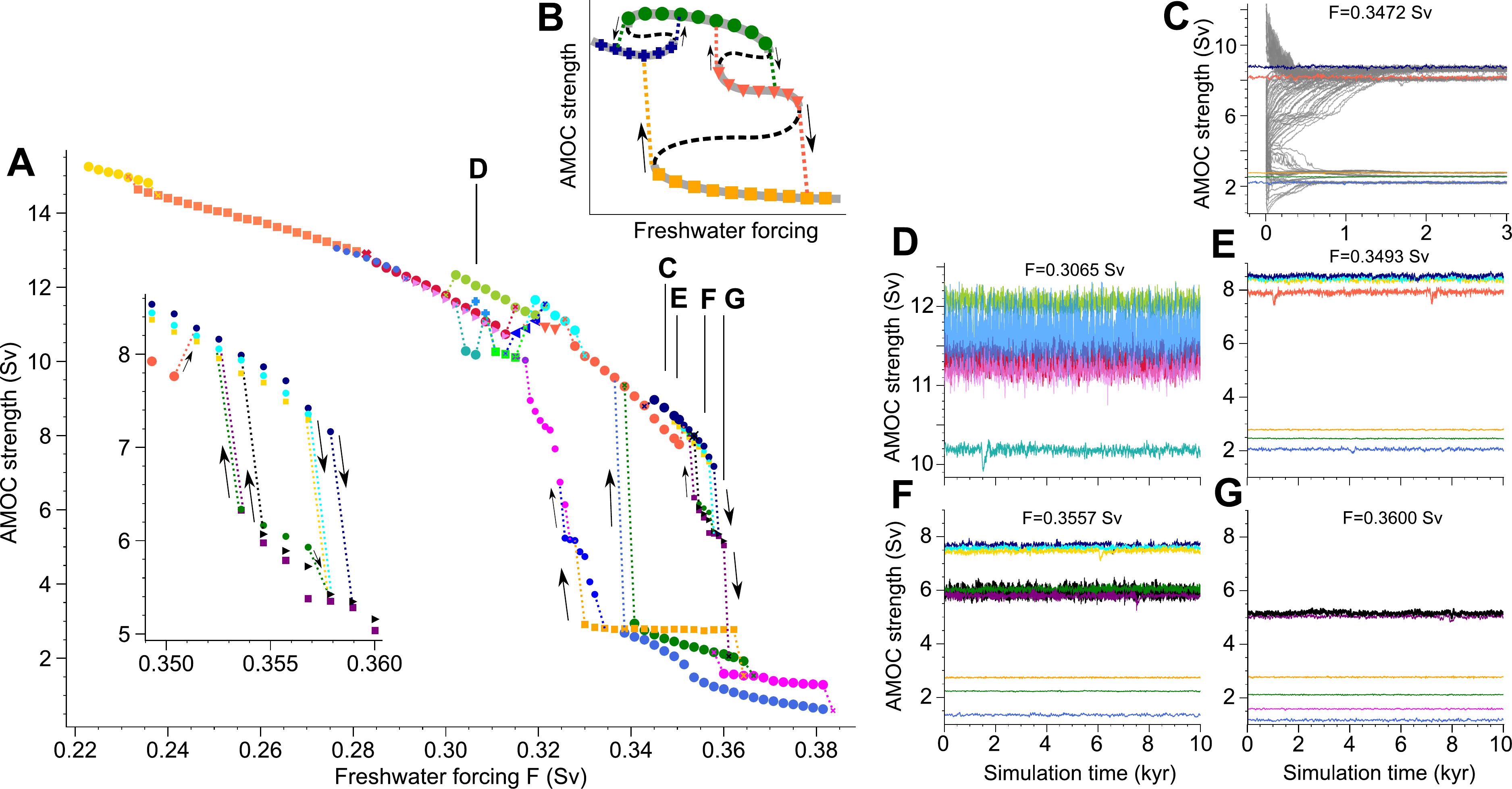}
\caption{\label{fig:multistability} 
{\bf Rich multistability of the AMOC obtained by a large ensemble of equilibrium simulations.}
{\bf A} Bifurcation diagram of the AMOC state with the freshwater forcing as control parameter. The y-axis shows the mean AMOC strength during the last 1000 years of each equilibrium simulation. The inset gives a detailed view of the coexisting attractors in the partially-collapsed and vigorous regimes shortly before the AMOC collapse. 
{\bf B} Schematic explainer of {\bf A}, using fabricated data. Symbols with the same color and shape correspond to the individual equilibrium simulations that represent one branch of stable AMOC states (thick gray lines). Coexisting branches have been reached from different initial conditions (Methods). The color of the dotted lines, and the arrows, indicate where a branch of attractors collapses to after loss of stability. Hypothetical unstable states lying between attractors are shown as black dashed lines.
{\bf C} Ensemble of N=89 simulations (gray) with perturbed initial conditions (Methods) that converge onto the different attractors (colored time series).
{\bf D-G} Long time series of the equilibrium simulations for all coexisting attractors at four values of the forcing parameter. The color coding corresponds to the branches in panel A.
}
\end{figure}

\begin{figure}%[floatfix]%!htb
\includegraphics[width=0.99\textwidth]{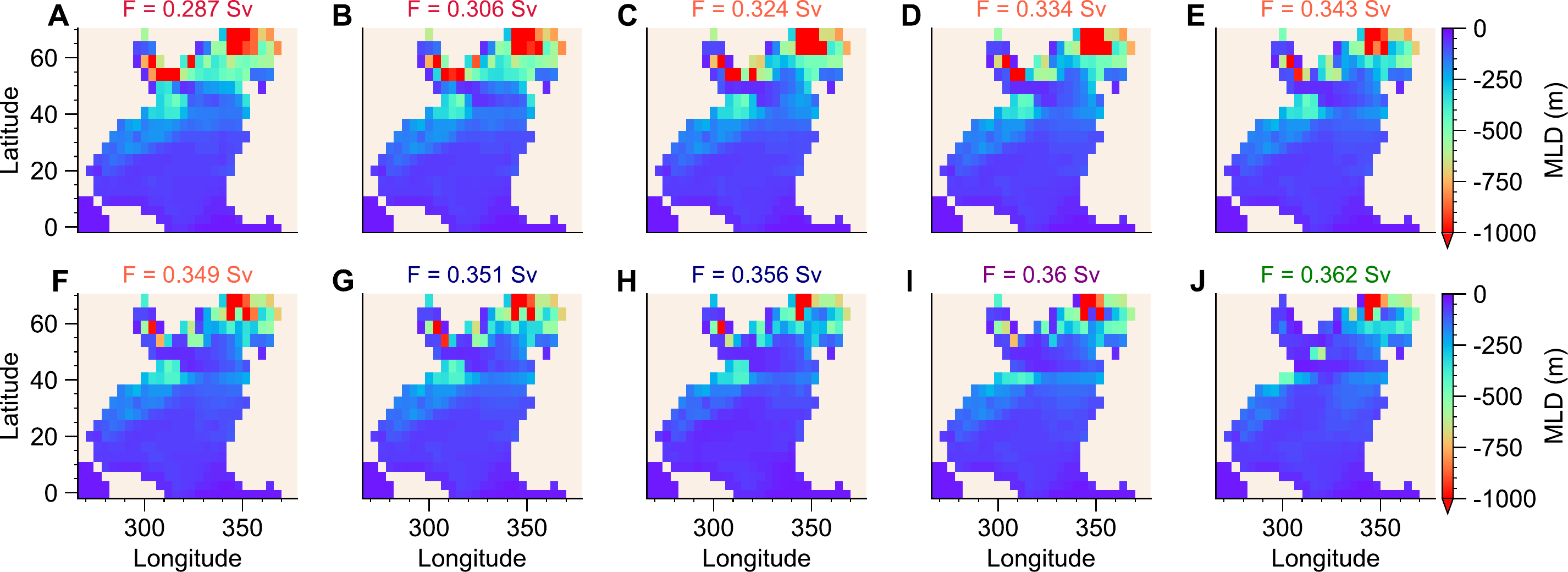}
\caption{\label{fig:mld_sequence} 
{\bf Change in spatial pattern of convection during the AMOC collapse.} Maps of the 100-year average winter mixed layer depth (MLD) in the NA for a series of 10 equilibrium simulations at increasingly larger freshwater forcing, given in the panel titles. MLD is defined by the depth until which the monthly average ocean temperature is within 0.5 K of the sea surface temperature, indicating the degree of vertical mixing via convection.
The winter MLD is obtained by averaging over the first four months of the year. The color coding in the panel titles indicates the branches of attractors in Fig.~3A. Panels {\bf A-H} are vigorous AMOC states, whereas {\bf I} is in the partially-collapsed regime, and {\bf J} is a collapsed AMOC state.
}
\end{figure}

\begin{figure}%[floatfix]%!htb
\includegraphics[width=0.99\textwidth]{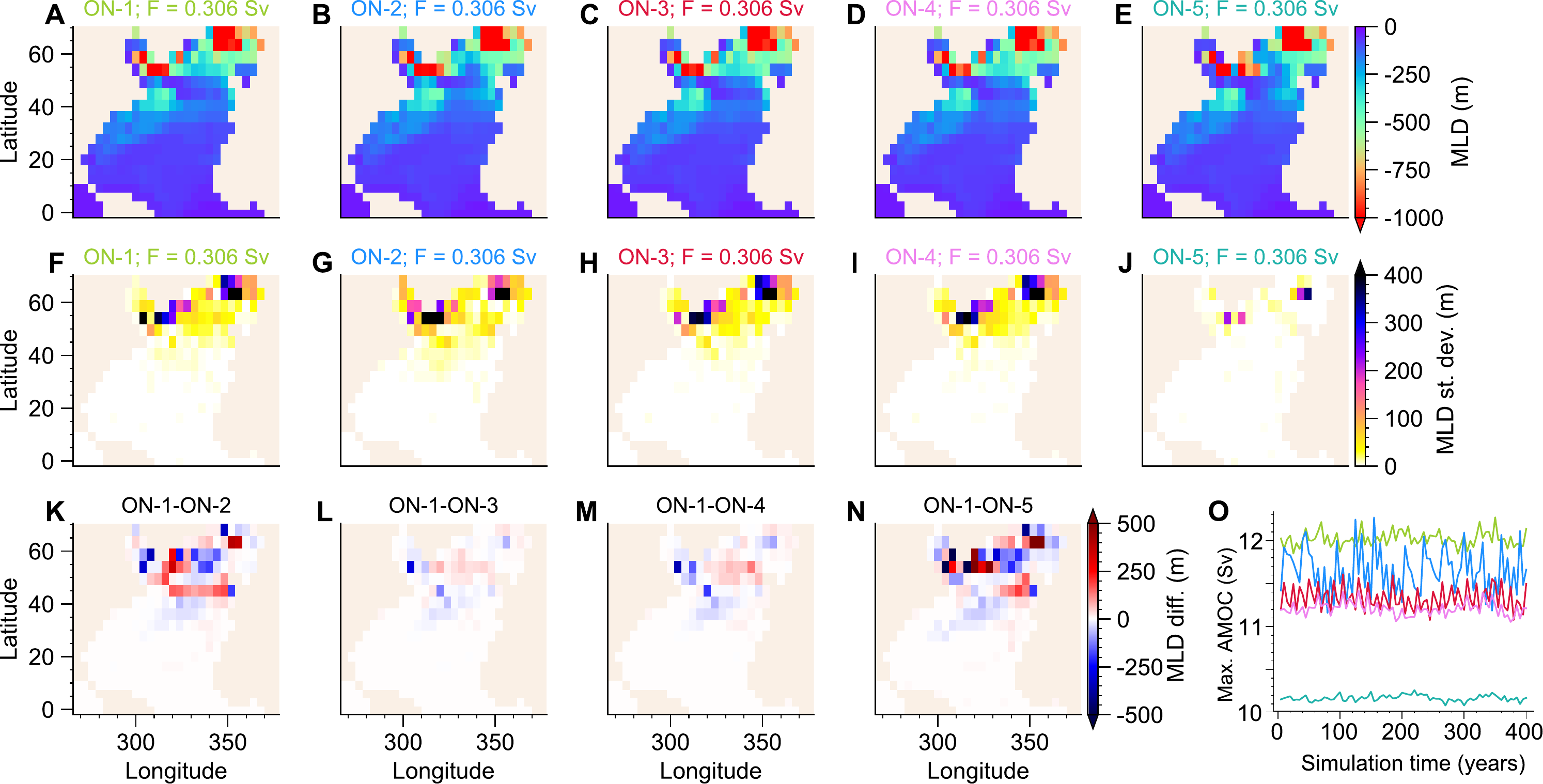}
\caption{\label{fig:mld_cluster} 
{\bf Spatial pattern and variability of convection in a cluster of coexisting stable states.}
{\bf A-E} Maps of the 100-year average MLD in the NA for the cluster of 5 coexisting vigorous AMOC states at the same forcing value (corresponding to the states shown in Fig.~3D). 
{\bf F-J} Temporal standard deviation of the MLD in consecutive winters obtained from 100 years of simulation. {\bf K-N} Difference in the 100-year mean MLD with respect to the state ON-1 of panel {\bf A}.
{\bf O} Short time series segments of the equilibrium simulations, highlighting the differences in AMOC variability. The same color code is used as in panels {\bf A-E}, and in Fig.~3.
}
\end{figure}

\begin{figure}%[floatfix]%!htb
\includegraphics[width=0.85\textwidth]{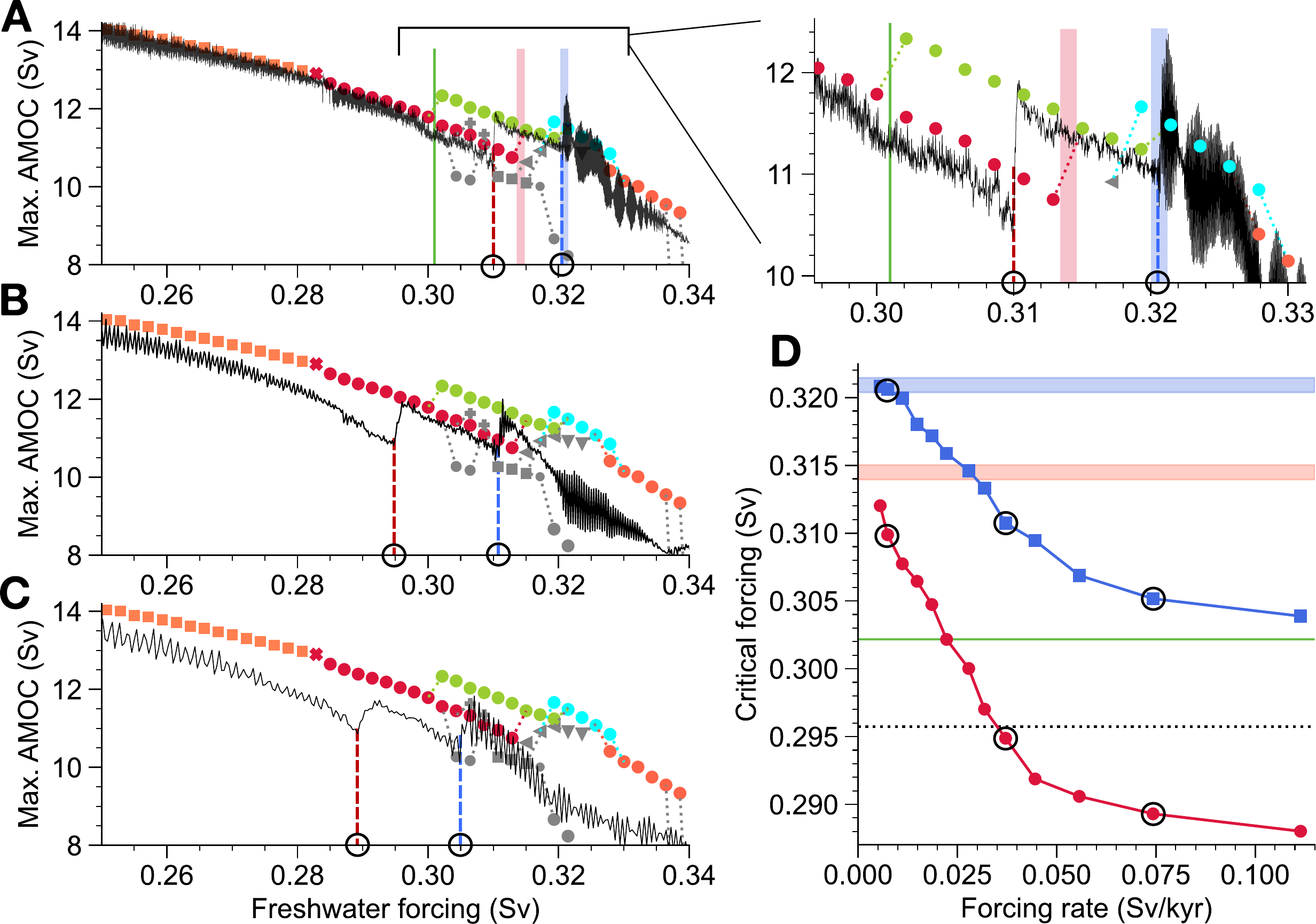}
\caption{\label{fig:transient_dynamics} 
{\bf Rate-dependence of intermediate tipping events.}
{\bf A-C} Transient simulations (black trajectories) with a linear forcing increase at different rates ({\bf A} 0.007 Sv/kyr, {\bf B} 0.037 kyr, and {\bf C} 0.074 Sv/kyr), superimposed on the upper portion of the bifurcation diagram of Fig.~3. The colored symbols are the branches of attractors that are visited by all transient trajectories, as has been determined by stopping the parameter shifts at several values. Other branches of attractors are given in gray, or omitted. Note, however, that for $F > 0.3$ Sv the basins of attractions of additional branches may be visited temporarily, depending on the rate (Fig.~S20). Dashed vertical lines indicate the forcing values at which the two most prominent ITPs occur. The thick red (blue) vertical line indicates the forcing value where the red (light green) branch loses stability. 
The green vertical line indicates the lowest forcing value where the light green branch exists. {\bf D} Forcing values where the two prominent ITPs (marked in panels {\bf A-C}) occur, as a function of the forcing rate of change during the linear parameter shift. 
}
\end{figure}

\begin{figure}%[floatfix]%!htb
\includegraphics[width=0.6\textwidth]{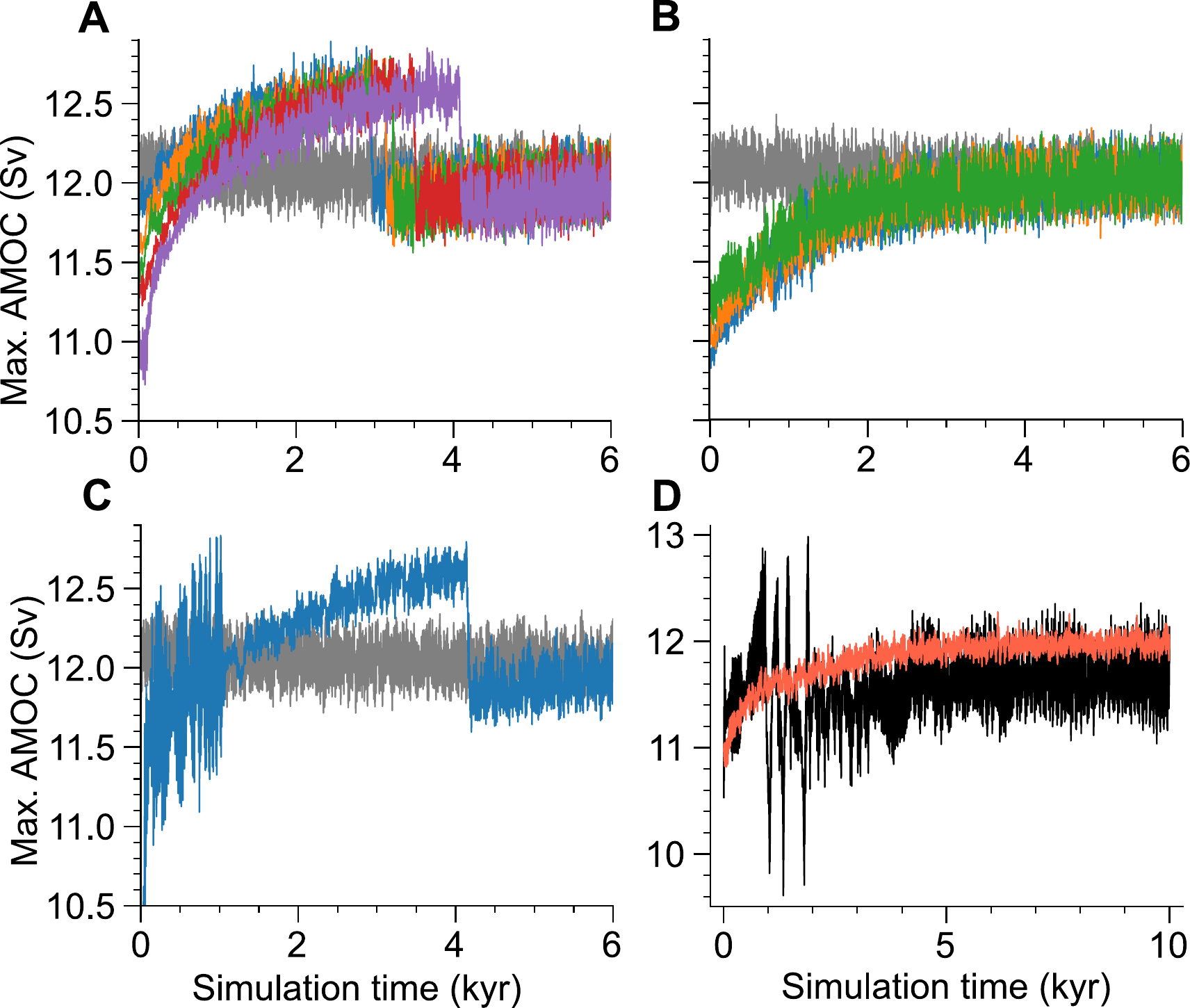}
\caption{\label{fig:transient_tipping} 
{\bf Transient tipping for moderately fast changes in freshwater forcing.}
{\bf A-C} Time series of simulations branched off at constant forcing value $F=0.296 \, \text{Sv}$ (dashed line in Fig.~6D) from the ensemble of linear ramping experiments (see main text). The simulations are branched off from ensemble members with ramping rates from ({\bf A}) 0.037 Sv/kyr to 0.111 Sv/kyr, ({\bf B}) 0.022 Sv/kyr to 0.032 Sv/kyr, and ({\bf C}) 0.223 Sv/kyr. The gray trajectory is the equilibrium simulation at the corresponding forcing value (red branch in Fig.~3).
{\bf D} Two constant-forcing simulations at $F=0.307 \, \text{Sv}$ branched off from the linear ramping experiments, where this forcing value was reached with the different forcing rates of 0.055 Sv/kyr (black) and 0.037 Sv/kyr (red).
}
\end{figure}

\begin{figure}%[floatfix]%!htb
\includegraphics[width=0.85\textwidth]{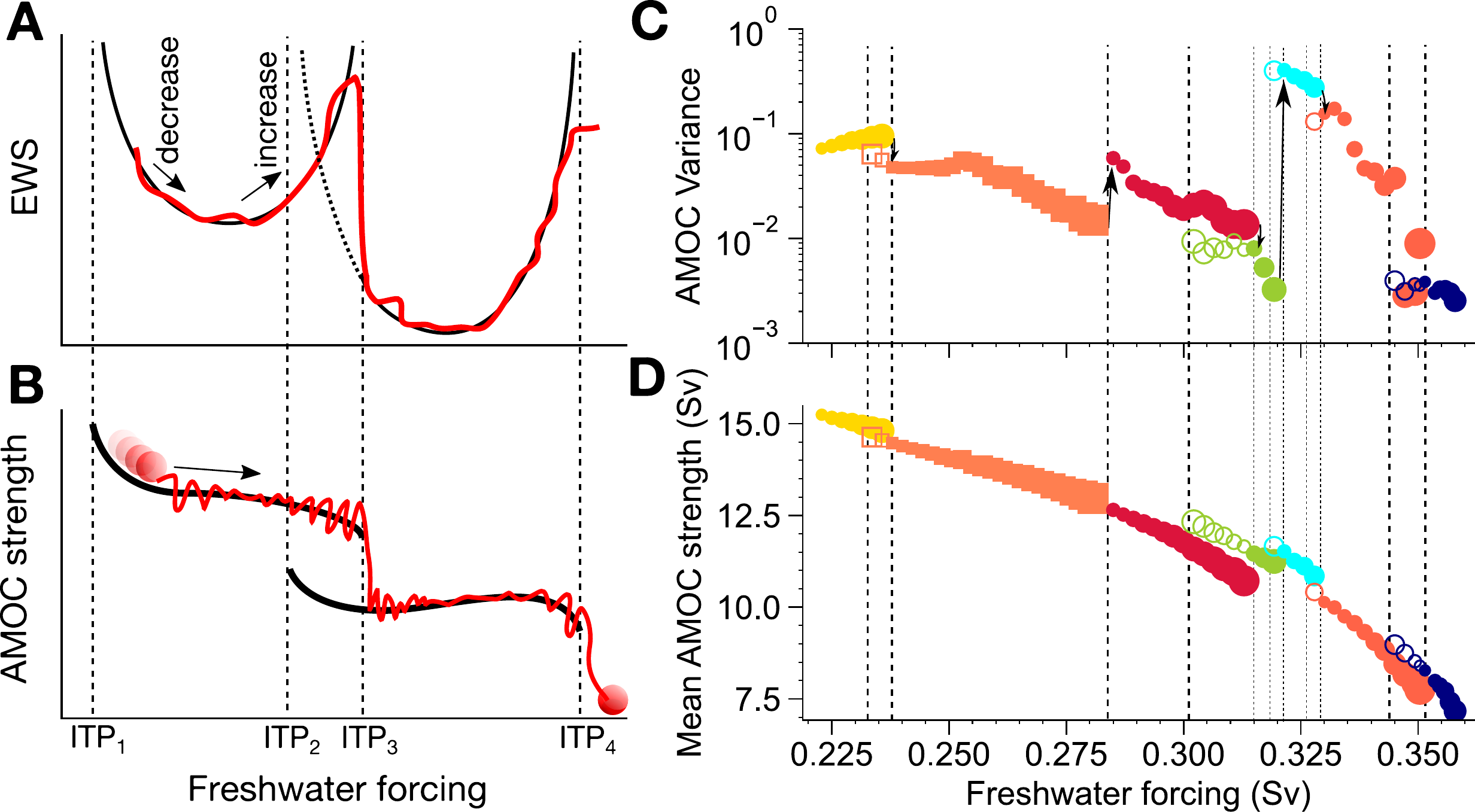}
\caption{\label{fig:ews_path} 
{\bf Changes in variability along the attractor branches leading up to the AMOC collapse.} 
{\bf A,B} Conceptual behavior of an ideal EWS indicator ({\bf A}) in a sequence of two attractor branches ({\bf B}) and corresponding ITPs. In a given branch, the EWS increases at the two ITPs (vertical dashed lines) for low and high forcing, yielding a u-shaped function (black curves in {\bf A}). A monotonic forcing increase across ITPs (red trajectories) can thus lead to non-monotonic trends, as well as jumps in the EWS.
{\bf C,D} Corresponding behavior of the variance as EWS in our model simulations.
{\bf C} Standard deviation as a function of freshwater forcing, calculated from the last 1500 years of all equilibrium simulations comprising the sequence of branches and states that is obtained when following the bifurcation diagram (Fig.~3A) from the top left until the last vigorous state at around $F=0.36 \, \text{Sv}$. The color coding is as in Fig.~3A, and the mean AMOC strength is shown in {\bf D}. Filled symbols are the parts of the branches that are actually visited during a monotonic freshwater increase. The open symbols are the other parts of the branches that are reached when instead decreasing the freshwater forcing.
}
\end{figure}

\begin{figure}%[floatfix]%!htb
\includegraphics[width=0.85\textwidth]{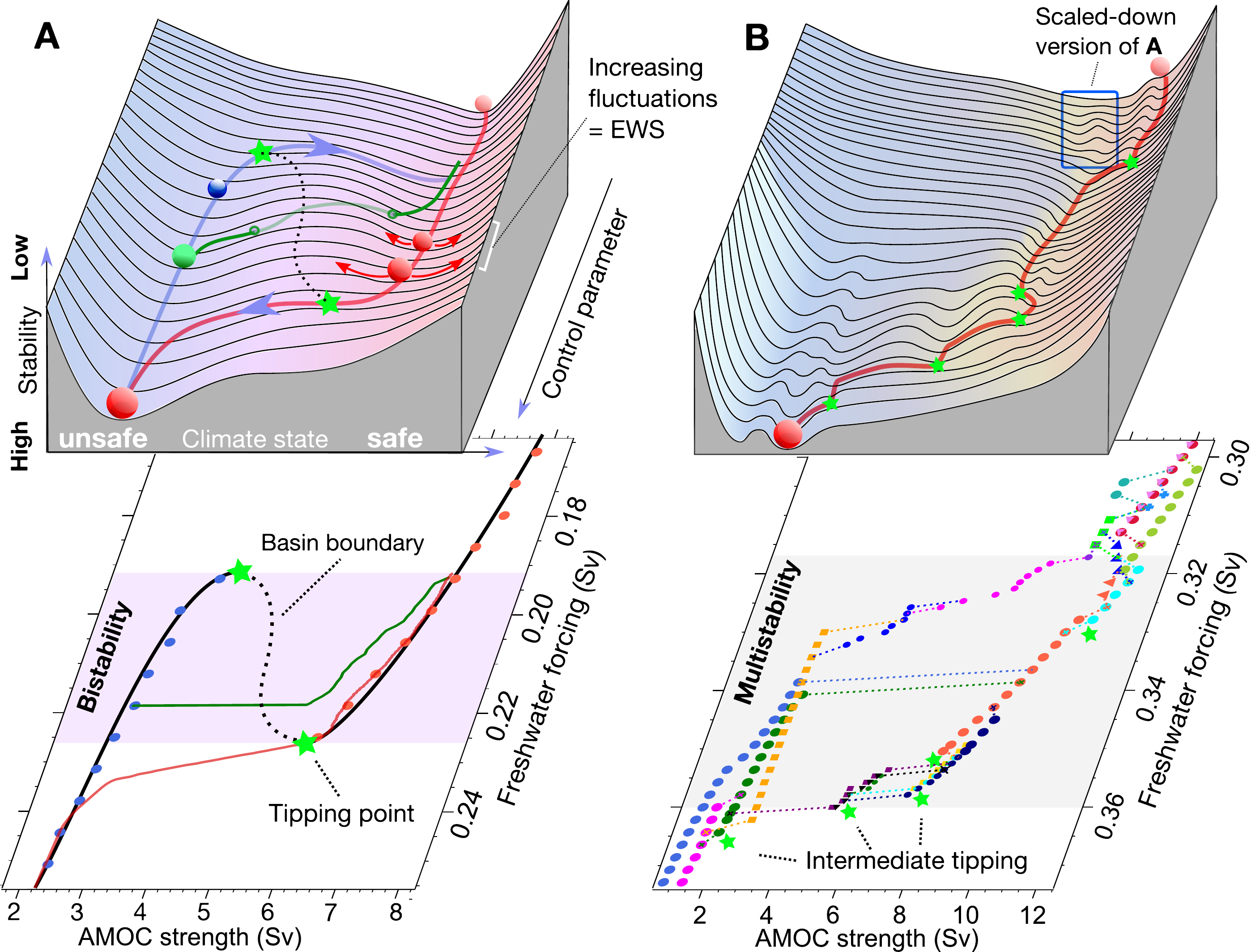}
\caption{\label{fig:stability_landscapes} 
{\bf Contrasting stability landscapes and tipping trajectories of bistable versus highly multistable systems.}
{\bf A} Bistable stability landscape (top) superimposed on our previously published simulations of an AMOC collapse \cite{LOH21}. For slow rates of forcing increase, the system's trajectory (red) closely follows the stable equilibrium states (valleys in the landscape above, colored symbols below), until a TP is reached, which may be preceded by EWS. 
For fast forcing (green trajectory) rate-induced tipping can occur before the TP, where the system is not able to follow its equilibrium \cite{LOH21}. 
{\bf B} Rugged stability landscape heuristically derived from the simulations presented here. Up to nine stable states coexist in different regimes and disappear in ITPs (green stars). This makes tipping more gradual, yet stepwise, dependent on initial conditions, and the task of early-warning highly non-trivial. 
}
\end{figure}

\end{document}